\documentclass[10pt,twocolumn]{IEEEtran}
\usepackage{epsfig,epsf,color,amsmath,cite,epic,eepic}
\usepackage{wrapfig}
\usepackage{bm}
\usepackage{amssymb}
\usepackage{url}
\usepackage[framemethod=tikz]{mdframed}

\allowdisplaybreaks[3]

\input{mysymbol.sty}
\long\def\symbolfootnote[#1]#2{\begingroup
	\def\thefootnote{\fnsymbol{footnote}}
	\footnote[#1]{#2}\endgroup} \psfull

\begin{document}
	
\title{\huge Signal Processing on Directed Graphs}
	
\author{{\it Antonio G. Marques, Santiago Segarra, and Gonzalo Mateos$^\dag$}\vspace{-15pt}}
	
\maketitle \maketitle \symbolfootnote[0]{$\dag$ Work in this paper
	was supported by the Spanish Fed. Grants Klinilycs and SPGraph (TEC2016-75361-R, PID2019-105032GB), and the USA NSF awards CCF-1750428 and ECCS-1809356. A. G. Marques is with the Dept. of Signal Theory \& Comms., King Juan Carlos University. S. Segarra is with the Dept. of Electrical \& Computer Eng. (ECE), Rice University. G. Mateos is with the Dept. of ECE, University of Rochester.  Emails: \texttt{antonio.garcia.marques@urjc.es}, \texttt{segarra@rice.edu},  and \texttt{gmateosb@ece.rochester.edu.}}

\markboth{DRAFT: Accepted to be Published in the IEEE Signal Processing Magazine Special Issue on Graph Signal Processing}
\maketitle \maketitle

\vspace*{-10pt}
\begin{abstract}%
This paper provides an overview of the current landscape of signal processing (SP) on directed graphs (digraphs). Directionality is inherent to many real-world (information, transportation, biological) networks and it should play an integral role in processing and learning from network data. We thus lay out a comprehensive review of recent advances in SP on digraphs, offering insights through comparisons with results available for undirected graphs, discussing emerging directions, establishing links with related areas in machine learning and causal inference in statistics, as well as illustrating their practical relevance to timely applications. To this end, we begin by surveying (orthonormal) signal representations and their graph frequency interpretations based on novel measures of signal variation for digraphs. We then move on to filtering, a central component in deriving a comprehensive theory of SP on digraphs. Indeed, through the lens of filter-based generative signal models, we explore a unified framework to study inverse problems (e.g., sampling and deconvolution on networks), statistical analysis of random signals, and topology inference of digraphs from nodal observations.  
\end{abstract}

\vspace*{-2pt}

\begin{IEEEkeywords}
\noindent Digraphs, Graph signal processing, Machine learning over graphs, Graph Fourier transform, Directed graphical models.
\end{IEEEkeywords}

\section{Introduction and Motivation}\label{sec:intro}
Coping with the panoply of challenges found at the confluence of data and network sciences necessitates 
fundamental breakthroughs in modeling, identification, and controllability of networked (complex) system processes -- often conceptualized as signals defined on graphs~\cite{Ortega2018ProcIEEE}. Graph-supported signals abound in real-world applications, including vehicle congestion levels over road networks, neurological activity signals supported on brain connectivity networks, and fake news that diffuse on online social networks. There is, however, an evident mismatch between our scientific understanding of signals defined over regular domains such as time or space and graph signals, due, in part, to the fact that the prevalence of network-related problems and access to quality network data are recent events. To address these problems, machine learning and signal processing (SP) over graphs have emerged as active areas aimed at making sense of large-scale \textit{datasets from a network-centric perspective}. Upon modeling the domain of the information as a graph and the observations at hand as graph signals, the graph SP (GSP) body of work has put forth models that relate the properties of the signals with those of the graph, along with algorithms that fruitfully leverage this relational structure to better process and learn from network data. Most GSP efforts to date assume that the underlying networks are  \textit{undirected}~\cite{djuric2018cooperative}.  Said graphs are equivalently represented by \textit{symmetric matrices} whose (well-behaved) spectral properties can be used to process the signals associated with the network. The most prominent example is  the graph Laplacian, which not only gives rise to a natural definition of signal smoothness but also offers a complete set of orthonormal eigenvectors that serve as a Fourier-type basis for graph signals~\cite{EmergingFieldGSP}. 

Their scarcer adoption notwithstanding, directed graph (digraph) models are more adequate (and, in fact, more accurate) for a number of applications. Information networks such as scientific citations or the Web itself are typically directed, and flows in technological (e.g., transportation, power, communication) networks are oftentimes one-directional. The presence of directionality plays a critical role when the measurements taken in those networks need to be processed to remove noise, outliers and artifacts, and this requires new tools and algorithms that do not assume that the  matrices representing the underlying graphs are symmetric. Gene-regulatory networks are highly non-reciprocal and this lack of reciprocity needs to be accounted for when, for example, the goal is to predict a gene or a protein functionality from a small set of observations obtained from expensive experiments. Pairwise relations among social actors are rarely purely symmetric \cite{kolaczyk2009book} and, in fact, when the graph captures some level of influence on a social network, the lack of symmetry is essential to accurately solve inverse problems that aim to separate the leaders from the followers~\cite{segarra2017blind}. More abstractly, when the graph encodes (oftentimes unknown) relations between observed variables, directionality is vital to identify the nodes representing the cause and those representing the effect~\cite{Peters2017}, calling for fundamental changes in the algorithms that use available signal observations to learn the topology of the underlying graph. Accordingly, a first step to address these and other related questions is to develop judicious models that account for directionality, while leading to tractable processing tools and efficient algorithms. 

In this context, this tutorial article aims at delineating the analytical background and the relevance of innovative tools to analyze and process signals defined over digraphs. More concretely, we will start by discussing different generalizations of smoothness and total variation measures for signals defined on digraphs. Those will then be used as a starting point to define orthonormal transforms and dictionaries for graph signals that account for the directionality of the supporting graph, including different generalizations of the graph Fourier transform (GFT) for digraphs (Section \ref{S:GFT_and_Dics_for_Digraphs}). In Section \ref{S:Filters_for_Digraphs}, we will introduce graph filters and discuss how to leverage these linear operators (akin to convolutions) to build more general information processing transformations (including deep nonlinear architectures). Building on the aforementioned two pillars of SP for digraphs (the GFT and linear graph filters), we will shift gears to a number of more advanced data-analytic tasks along with GSP tools to address them. These include: (i)~network inverse problems such as sampling, deconvolution, and system identification (Section \ref{S:inverse_probs});  (ii)~statistical models for random graph signals over digraphs (Section \ref{S:Satistical_SP}); and (iii)~algorithms to identify the topology of directed graphs in Section \ref{S:topo_id}. In this last topic, connections with causality will be stressed and emerging problems will be identified. Throughout, concepts will be made accessible to SP researchers (including those without a strong background on network science) via a combination of rigorous problem formulations and intuitive reasoning. In Section \ref{S:Apps}, we present several illustrative applications involving real datasets to showcase the potential benefits of the tools previously discussed. A recurrent message with important practical ramifications interweaves the narrative -- different from the undirected case where graph spectrum-based tools offer a number of distinct advantages~\cite{EmergingFieldGSP}, vertex-domain generative graph-signal models that rely on non-symmetric network operators may be preferable when it comes to signal and information processing on directed networks. Emerging topics and open problems at the frontier of SP on digraphs are the subject of the concluding summary in Section \ref{S:Conclusions}.
%
 

\section{Graph signal processing preliminaries,  frequency analysis, and signal representations}\label{S:GFT_and_Dics_for_Digraphs}

After introducing the necessary graph-theoretic notation and background, this section will present different generalizations of smoothness and total variation measures for signals defined on digraphs. This is particularly relevant to the GFT, which decomposes a graph signal into  components describing different modes of variation with respect to the graph topology. While for \emph{undirected} graphs adopting the real-valued \textit{orthonormal} eigenvectors of the Laplacian as the frequency basis is well motivated and widely used in practice~\cite{Ortega2018ProcIEEE}, extending the GFT framework to digraphs is not a simple pursuit and different alternatives exist, as we explain in Sections \ref{Ss:GFT_for_Digraphs} and \ref{Ss:orthonormal_transforms}.

\vspace{.15cm} 
\noindent \textbf{Graph signals and the graph-shift operator.} Let $\mathcal{G}$ denote a \textit{directed} graph with a set of nodes $\mathcal{N}$ (with cardinality $N$) and a set of links $\mathcal{E}$, if $i$ is connected to $j$ then $(i,j)\in\mathcal{E}$. Since $\ccalG$ is directed, local connectivity is captured by the set $\ccalN_i:=\{j\;|(j,i)\in\mathcal{E}\}$ which stands for the (incoming) neighborhood of $i$. For any given $\mathcal{G}$ we define the adjacency matrix $\bbA\in\mathbb{R}^{N\times N}$ as a sparse matrix with non-zero elements $A_{ji}$ if and only if $(i,j)\in\ccalE$. The value of $A_{ji}$ captures the strength of the connection from $i$ to $j$ and, since the graph is directed, the matrix $\bbA$ is in general non-symmetric.

The focus of the paper is on analyzing and modeling (graph) signals defined on the node set $\mathcal{N}$. These signals can be represented as vectors $\mathbf{x}=[x_1,...,x_N]^T \in  \mathbb{R}^N$, with $x_i$ being the value of the signal at node $i$. Since the vectorial representation does not account explicitly for the structure of the graph,  $\mathcal{G}$ can be endowed with the so-called graph-shift operator (GSO) $\mathbf{S}$ \cite{SandryMouraSPG_TSP13,SandryMouraSPG_TSP14Freq}. The shift $\mathbf{S}\in\mathbb{R}^{N\times N}$ is a matrix whose entry $S_{ji}$ can be non-zero only if $i=j$ or if $(i,j)\in\mathcal{E}$. The sparsity pattern of the matrix $\bbS$ captures the local structure of $\ccalG$, but we make no specific assumptions on the values of its non-zero {entries, which will depend on the application at hand~\cite{Ortega2018ProcIEEE}}. To justify the adopted {\emph{graph shift}} terminology, consider the directed cycle graph whose circulant adjacency matrix $\bbA_{dc}$ is zero, except for entries $A_{ji}=1$ whenever $i=\mathrm{mod}_N(j)+1$, where $\mathrm{mod}_N(x)$ denotes the modulus (remainder) obtained after dividing $x$ by $N$. Such a graph can be used to represent the domain of discrete-time periodic signals with period $N$. If $\bbS=\bbA_{dc}$, then $\bbS\bbx$ implements a {\emph{circular shift}} of the entries in $\bbx$, which corresponds to a one-unit time delay under the aforementioned interpretation~\cite{Ortega2018ProcIEEE}.  Notice though that in general $\bbS$ need be neither invertible nor isometric, an important departure from the shift in discrete-time SP. The intuition behind $\mathbf{S}$ is to represent a linear transformation that can be computed locally at the nodes of the graph, while it can be more general than the adjacency matrix. More rigorously, if the graph signal $\mathbf{y}$ is defined as $\mathbf{y}=\mathbf{S}\mathbf{x}$, then node $i$ can compute $y_i$ as a linear combination of the signal values $x_j$ at node $i$'s neighbors $j\in \mathcal{N}_i$. The GSO will play a fundamental role in defining the counterpart of the Fourier transform for graph signals, which is discussed in this section, as well as graph filters that are introduced in Section \ref{S:Filters_for_Digraphs}. 

{\subsection{Digraph Fourier transforms: Spectral methods}\label{Ss:GFT_for_Digraphs}}

An instrumental GSP tool is the GFT, which decomposes a graph signal into orthonormal components describing different modes of variation with respect to the graph topology encoded in  an application-dictated GSO $\bbS$. The GFT allows to equivalently represent a graph signal in two different domains -- the vertex domain consisting of the nodes in $\ccalN$, and the graph frequency domain spanned by the spectral basis of \ccalG. Therefore, signals can be manipulated in the frequency domain for the purpose of e.g., denoising, compression, and feature extraction; see also Section \ref{S:Filters_for_Digraphs}. For didactic purposes it is informative to introduce first the GFT for symmetric graph Laplacians associated with undirected graphs; see the callout \emph{``A motivating starting point: The GFT for undirected graphs''}. In the remainder of this section, we show that the GFT can be defined for digraphs where the interpretation of components as different modes of variability is not as clean and Parseval's identity may not hold, but its value towards yielding parsimonious spectral representations of network processes remains. 

\begin{mdframed}[hidealllines=true,backgroundcolor=gray!20]
\textbf{A motivating starting point: The GFT for undirected graphs.} Consider an undirected graph {\ccalG} with combinatorial Laplacian $\bbL=\bbD-\bbA$ chosen as GSO~\cite{EmergingFieldGSP}, where $\bbD$ stands for the diagonal degree matrix. The symmetric $\bbL$ can always be decomposed as $\bbL =  \bbV \diag(\bblambda) \bbV^T$, with $\bbV := [\bbv_1,...,\bbv_{N}]$ collecting the orthonormal eigenvectors of the Laplacian and $\bblambda:=[\lam_1,...,\lam_{N}]^T$ its non-negative eigenvalues. The GFT of $\bbx$ with respect to $\bbL$ is the signal $\tbx=[\tdx_1,...,\tdx_{N}]^T$ defined as $\tbx = \bbV^T\bbx.$ The inverse GFT (iGFT) of $\tbx$ is given by $\bbx = \bbV \tbx$, which is a proper inverse by the orthogonality of $\bbV$.
	
The iGFT formula $\bbx = \bbV \tbx=\sum_{k=1}^{N} \tdx_k \bbv_k$ allows one to synthesize $\bbx$ as a sum of orthogonal frequency components $\bbv_k$. The contribution of $\bbv_k$ to the signal $\bbx$ is the real-valued GFT coefficient $\tdx_k$. The GFT encodes a notion of signal variability over the graph akin to the notion of frequency in Fourier analysis of temporal signals. To understand this analogy, define the total variation of the graph signal $\bbx$ with respect to the Laplacian $\bbL$ (also known as Dirichlet energy) as the following quadratic form
\begin{align}\label{eqn_TV_general}
\text{TV}_2(\bbx) := \bbx ^T \bbL \bbx  = \sum_{i < j} A_{ij} (x_i-x_j)^2.
\end{align}
The total variation $\text{TV}_2(\bbx)$ is a smoothness measure, quantifying how much the signal $\bbx$ changes with respect to the {graph topology encoded} in $\bbA$. 
	
Back to the GFT, consider the total variation of the eigenvectors $\bbv_k$, which is given by $ \text{TV}_2(\bbv_k) = \bbv_k^T \bbL \bbv_k = \lam_k$. It follows that the eigenvalues $0=\lam_1 \leq \lam_2\leq\ldots\leq\lam_{N}$ can be viewed as graph frequencies, indicating how the eigenvectors (i.e., frequency components) vary over the graph $\ccalG$. Accordingly, the GFT and iGFT offer a decomposition of the graph signal $\bbx$ into spectral components that characterize different levels of variability. 
\end{mdframed}

{The Laplacian $\bbL=\bbD-\bbA$ is not well defined for digraphs because $\bbD$ is rendered meaningless when edges have directionality. One can instead consider a generic asymmetric GSO $\bbS$, for instance  the adjacency matrix $\bbA$ or one of the several generalized Laplacians for digraphs; see e.g., ~\cite{sevi2018harmonic,chung2005laplacians}. Suppose the GSO is diagonalizable as $\bbS=\bbV\diag(\bblambda) \bbV^{-1}$, with $\bbV := [\bbv_1,...,\bbv_{N}]$ denoting the (non-orthogonal) eigenvectors  of $\bbS$ and $\bblambda:=[\lam_1,...,\lam_{N}]^T$ its possibly complex-valued eigenvalues.}  Then a widely-adopted alternative is to redefine the GFT as $\tbx = \bbV^{-1}\bbx$~\cite{SandryMouraSPG_TSP14Freq}. {Otherwise, one can resort to the  Jordan decomposition of $\bbS$ and use its generalized eigenvectors as the GFT basis; see also~\cite{deri2017spectral} for a careful treatment of the non-diagonalizable case which relies on oblique spectral projectors to define the GFT.} Setting the GFT to $\bbV^{-1}$ for the directed case is an intuitively pleasing definition, since frequency components correspond to the eigenvectors of a (reference) shift operator as in discrete-time SP. 
Moreover, allowing for generic GSOs reveals the encompassing nature of the GFT relative to the time domain discrete Fourier transform (DFT), the multidimensional DFT, and Principal Component Analysis (PCA)~\cite{SI_SPMAG2019}. 
Towards interpreting graph frequencies which are defined by the (possibly complex-valued, non-orthogonal) eigenvectors of the non-symmetric $\bbS$, consider the total variation measure
\begin{align}\label{eqn_TV_directed}
\text{TV}_1(\bbx) := \|\bbx-\bar{\bbS}\bbx\|_1,
\end{align}
where $\bar{\bbS}=\bbS/|\lambda_{\max}|$ and $\lambda_{\max}$ is the spectral radius of $\bbS$ [cf. \eqref{eqn_TV_general}]. Using \eqref{eqn_TV_directed} and following the rationale for undirected graphs, one can define a frequency ordering $\lambda_i\succ\lambda_j$ if $\text{TV}_1(\bbv_i) >\text{TV}_1(\bbv_j) $~\cite{SandryMouraSPG_TSP14Freq}. While applicable to signals on digraphs, unlike \eqref{eqn_TV_general} the  signal variation measure \eqref{eqn_TV_directed} does not ensure that constant signals have zero variation. In addition, (generalized) eigenvectors of \emph{asymmetric} GSOs need not be orthonormal, implying that Parseval's identity will not hold {and hence the signal power is not preserved across the vertex and dual domains.}  {This in general can be an issue for graph filtering methods operating in the spectral domain, thus, motivating this paper's overarching theme of relying on vertex domain operations for extensions to digraphs.} {From a computational standpoint}, obtaining the Jordan decomposition for moderate-sized graphs is expensive and {often} numerically unstable; see also~\cite{deri2017spectral} and references therein for recent attempts towards mitigating this instability issue. Addressing uniqueness of the representation is also critical when the GSO (even the combinatorial Laplacian) has repeated eigenvalues, since the corresponding eigenspaces exhibit rotational ambiguities which can hinder interpretability of graph frequency analyses. To address this (often overlooked shortcoming),~\cite{deri2017spectral} puts forth a quasi-coordinate free GFT definition based on oblique spectral projectors. Other noteworthy GFT approaches rely on projections onto the (non-orthogonal) eigenvectors of a judicious random walk operator on the digraph~\cite{sevi2018harmonic,chung2005laplacians}; the interested reader is referred to~\cite[Section 7]{sevi2018harmonic} for a nice collection of examples involving semi-supervised learning and signal modeling on digraphs.

{Alternatives to the spectral GFT methods described so far are surveyed in the following section. 
The focus shifts to orthonormal transform learning approaches, whereby optimization problems are formulated to find suitable spectral representation bases for graph signals.}

{\subsection{Digraph Fourier transforms:  Orthonormal transform learning}\label{Ss:orthonormal_transforms}}

The history of SP has repeatedly taught us how low frequencies are more meaningful in human speech for the purpose of compression, high frequencies represent borders in images whose identification is key for segmentation, and different principal components offer varying discriminative powers when it comes to face recognition. While analogous interpretations are not always possible in more advanced representations obtained with modern tools such as learned overcomplete dictionaries and neural networks, at a basic level it remains true that orthonormal linear transformations
excel at separating signals from noise. Motivated by this general signal representation principle, a fresh look at the GFT for digraphs was put forth in \cite{sardellitti2017digraphFT} based on the minimization of {the} convex Lov\'{a}sz extension of the graph cut size {(which can be interpreted as a measure of signal variation on the graph capturing the edges' directionality)}, subject to \emph{orthonormality} constraints on the desired bases. The rationale behind the graph cut criterion is that its minimization leads to identifying clusters in $\ccalG$. 
Accordingly, the learned GFT bases in \cite{sardellitti2017digraphFT} tend to be constant across clusters of the graph, offering parsimonious spectral representations of signals that are real-valued and piecewise-constant over said clusters. The price paid for all these desirable properties is that the resulting GFT basis may fail to yield {atoms capturing different levels of signal variation with respect to $\ccalG$}, and the optimization procedure in~\cite{sardellitti2017digraphFT} is computationally expensive due to repeated singular value decompositions. 

A related (optimization-based) approach in~\cite{RAGE_GFT_TSP_2019} searches for an orthonormal digraph Fourier transform (DGFT) basis $\bbU := [\bbu_1, \ldots, \bbu_N] \in \reals^{N \times N}$, where $\bbu_k \in \reals^N$ represents the $k$th frequency component. Towards defining frequencies, a more general notion of signal directed variation (DV) for digraphs is introduced
%
$\text{DV}(\bbx) := \sum_{i\neq j}A_{ji} [x_i - x_j]_{+}^2$,
%
where $[x]_{+} := \max(0,x)$ denotes projection onto the non-negative reals. To gain insights on DV, consider a graph signal $\bbx$ on the digraph $\ccalG$ {and suppose a} directed edge represents the direction of signal flow from a larger value to a smaller one. Thus, an edge from node $i$ to node $j$ (i.e., $A_{ji}>0$) contributes to $\text{DV}(\bbx)$ only if $x_i > x_j$. Moreover, notice that if $\ccalG$ is undirected, then $\text{DV}(\bbx)\equiv\text{TV}_2(\bbx)$. In analogy to the GFTs surveyed in Section \ref{Ss:GFT_for_Digraphs}, we define the frequency $f_k := \text{DV}(\bbu_k)$ as the directed variation of the  frequency component $\bbu_k$. Since for all previous GFT approaches the spacing between frequencies can be highly irregular, the idea in~\cite{RAGE_GFT_TSP_2019} to better capture low, bandpass, and high frequencies is to design a DGFT such that the orthonormal frequency components are as spread as possible in the graph spectral domain.  Beyond offering parsimonious representations of slowly-varying signals on digraphs, a DGFT with spread frequency components can facilitate more interpretable frequency analyses and aid filter design in the spectral domain.  To this end,  a viable approach is to minimize a so-termed spectral dispersion criterion
\begin{align}
\label{eq:delta_opt_prob}
\bbU^* &= \argmin_{\bbU}
\:\sum_{i=1}^{N-1} \left[\text{DV}(\bbu_{i+1})-\text{DV}(\bbu_i)\right]^2 \quad\\ 
&\text{s. to}
\quad  \bbU^T\bbU=\bbI_N,\:\: \bbu_1=\frac{\mathbf{1}_{N}}{\sqrt{N}}, \:\:\bbu_N=\argmax_{\lVert \bbu \rVert = 1} \: \text{DV}(\bbu).\nonumber
\end{align}
The cost function measures how well spread the corresponding frequencies are over $[0,\textrm{DV}(\bbu_N)]$. Having fixed the first and last columns of $\bbU$, the dispersion function is minimized when the free directed variation values are selected to form
an arithmetic sequence over the attainable bandwidth.  
However, since the variables here are the columns of $\bbU$, we can only expect to obtain approximately equidistributed frequencies. Finding the global optimum of \eqref{eq:delta_opt_prob} is challenging due to the  non-convexity arising from the orthonormality (Stiefel manifold) constraints, a yet a stationary point can be provably obtained via the algorithm in~\cite{RAGE_GFT_TSP_2019}. Accordingly, the basis $\bbU^*$ in \eqref{eq:delta_opt_prob} and its counterpart in \cite{sardellitti2017digraphFT} may not be unique. 
In Section \ref{S:Apps}, we illustrate a graph-signal denoising task whereby the DGFT basis learned from \eqref{eq:delta_opt_prob} is used to decompose and then (low-pass) filter temperatures recorded across the United States.

\section{Graph filters and nonlinear graph-signal operators }\label{S:Filters_for_Digraphs}

\begin{figure}
	\centering\includegraphics[width=0.99\linewidth]{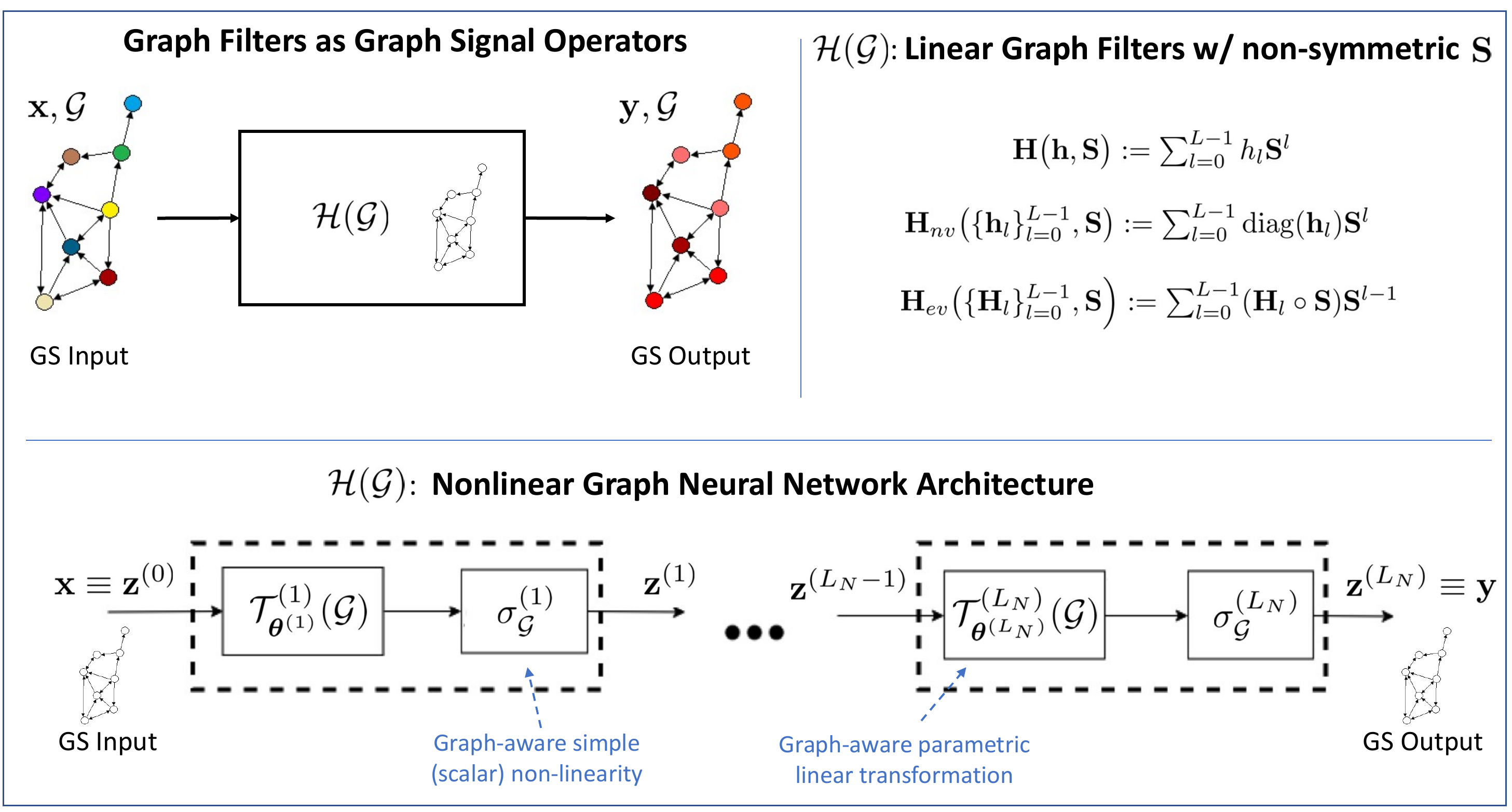}
	\vspace{-.3cm}
	\caption{(Top left) 
		Graph filters as generic operators that transform a (graph signal) input into a (graph signal) output. The graph filter processes the features of the input taking into account the topology of the digraph where the signals are defined. (Top right) Different types of linear graph filters: regular (shift-invariant) graph filter $\bbH$, node-variant graph filter $\mathbf{H}_{nv}$, and edge-variant graph filter $\mathbf{H}_{ev}$. The number of parameters (coefficients) is $L$, $NL$, and $|\ccalE| L$, respectively. Due to their polynomial definition, all these filters can operate over directed graphs (non-symmetric $\bbS$). (Bottom) Nonlinear graph signal operators using a (potentially deep) neural network with $L_N$ layers. Each layer consists of a parametrized graph-aware linear transformation (given, e.g., by any of the linear graph filters described before) followed by a point-wise nonlinearity [cf. \eqref{E:Generic_Graph_NN_architecture_input}-\eqref{E:Generic_Graph_NN_architecture_output}].}
	\vspace{-.2cm}
	\label{F:GraphFilters}
\end{figure}

Here we consider operators whose inputs and outputs are signals defined on a digraph; see the top-left panel of Fig.~\ref{F:GraphFilters} for a pictorial representation. These operators are not only used to process information defined on digraphs (see also the applications in Section \ref{S:Apps}), but also  to postulate (generative) signal models for network data and solve statistical inference tasks surveyed in Sections \ref{S:inverse_probs}-\ref{S:topo_id}. A key aspect throughout the discussion is how the topology of the digraph impacts the transformation of signals.  The section begins by discussing linear graph filters~\cite[Ch. 11]{djuric2018cooperative} and then builds on those to describe nonlinear (deep) architectures. After a brief outline of the current filtering landscape for undirected graphs, we will focus on recent progress to tackle the challenges faced when extending those operators to the directed case.

\subsection{Linear graph filters}
Several definitions for graph filters coexist in the GSP literature. Early works focused on using the graph Laplacian $\bbL$ as the GSO and leveraged its eigendecomposition $\bbL=\bbV \diag(\bblambda) \bbV^T$ (see the callout in Section \ref{Ss:GFT_for_Digraphs}) to define the graph filtering operation in the \textit{spectral} domain \cite{EmergingFieldGSP}. Specifically, if $\bbx$ denotes the input of the graph filter and $\bby$ its output, filtering a graph signal is tantamount to transforming the input signal to the graph Fourier domain as $\tbx=\bbV^T\bbx$, applying a \textit{point-wise} (diagonal) operator in the spectral domain to generate the output $\tby$, and finally transforming the obtained output back onto the vertex domain as $\bby=\bbV \tby$. The point-wise spectral operator can be expressed as the multiplication by a diagonal matrix $\diag(\tbg)$, so that $\tby=\diag(\tbg) \tbx$. Alternatively, one can adopt a scalar kernel function $g: \reals \rightarrow \reals$ applied to the eigenvalues of the Laplacian, so that the frequency response of the filter can be obtained as $\diag(\tbg)=\diag(g(\bblambda))$, where $g(\cdot)$ is applied entry-wise. Regardless of the particular choice, the input-output relation can be written as 
\begin{eqnarray}\label{E:Filter_input_output_frequency}
\bby= \bbV \diag(g(\bblambda)) \bbV^T \bbx= \bbV \diag(\tbg)\bbV^T \bbx,
\end{eqnarray} 
with the $N\times N$ matrix $\bbV \diag(g(\bblambda)) \bbV^T $ representing the linear transformation in the nodal domain. An alternative definition consists in leveraging the interpretation of $\bbS$ as a reference graph-signal operator and then building more general linear \emph{operators} of the form~\cite{SandryMouraSPG_TSP13}
\begin{eqnarray}\label{E:Filter_input_output_vertex}
\bby&=&h_0 \bbx + h_1 \bbS \bbx + ... + h_{L-1} \bbS^{L-1}\bbx := \bbH\bbx,\;\;\\
&~&\textrm{with}\;\;\mathbf{H}:=\sum_{l=0}^{L-1}h_l \mathbf{S}^l, \nonumber
\end{eqnarray}
where the filter coefficients are collected in $\mathbf{h}:=[h_0,\ldots,h_{L-1}]^T$, with $L-1$ denoting the filter degree. Upon defining $\bbx^{(l+1)}:=\bbS\bbx^{(l)}$ and $\bbx^{(0)}=\bbx$, the output $\bby$ in \eqref{E:Filter_input_output_vertex} can be equivalently written as $\bby=\sum_{l=0}^{L-1}h_l \bbx^{(l)}$. Since the application of the GSO $\bbS$ requires only local exchanges among (one-hop) neighbors, the latter expression reveals that the operators in \eqref{E:Filter_input_output_vertex} can be implemented in a distributed fashion with $L-1$ successive exchanges of information among neighbors~\cite{segarra2017optimal}. This is a key insight (and advantage) of \eqref{E:Filter_input_output_vertex} that will be leveraged in subsequent sections. {Note that the coefficients $\bbh$ can be given (e.g., when modeling a known network diffusion dynamics) or designed to accomplish a particular SP task such as low-pass filtering; see, e.g., \cite{segarra2017optimal} for further details on graph-filter designs.} 

When $L=N$ and $\ccalG$ is symmetric so that the GSO is guaranteed to be diagonalizable, the two previous definitions can be rendered equivalent. But this is not the case when the graph filter is defined on a digraph. To see why, note that the polynomial definition in \eqref{E:Filter_input_output_vertex} is valid regardless of whether the GSO is symmetric or not. Its interpretation as a local operator also holds true for digraphs, provided that the notion of \textit{locality} is understood in this case considering only the neighbors with \textit{incoming} connections. The generalization of the definition in \eqref{E:Filter_input_output_frequency} to the directed case is, however, more intricate. As explained in Section \ref{S:GFT_and_Dics_for_Digraphs}, different GFTs for digraphs exist. If the iGFT is given by the eigenvectors of the GSO, then one only needs to replace $\bbV^T$ with the (non-orthogonal) $\bbV^{-1}$ in \eqref{E:Filter_input_output_frequency}. If the GSO is diagonalizable and $\bbV^{-1}$ is adopted as the GFT, then the polynomial definition in \eqref{E:Filter_input_output_vertex} and the updated version of \eqref{E:Filter_input_output_frequency} are equivalent. If the GSO is not diagonalizable the generalization of \eqref{E:Filter_input_output_frequency} is unclear, while \eqref{E:Filter_input_output_vertex} still holds. On the other hand, if the GFT is not chosen to be $\bbV^{-1}$ but one of the orthogonal (graph-smoothness related) dictionaries $\bbU$ presented in Section \ref{Ss:orthonormal_transforms}, then the two definitions diverge. Specifically, linear operators of the form $\bbU \diag(\tbg) \bbU^T$ will be symmetric (meaning that the influence of the input at node $i$ on the output at node $j$ will be the same than that of node $j$ on node $i$), while operators of the form $\sum_{l=0}^{L-1} h_l \bbS^l$ will not. Equally important, while a polynomial filter can always be implemented using local exchanges, there is no guarantee that the symmetric transformation $\bbU \diag(\tbg) \bbU^T$ can be implemented in a distributed fashion~\cite{segarra2017optimal}. All in all, if the definition in \eqref{E:Filter_input_output_vertex} is adopted for the directed case then graph filters are always well defined, their distributed implementation is still feasible, and the design and interpretation of the filter coefficients $\bbh$ as weights given to the information obtained after successive local exchanges is preserved. Their interpretation as diagonal spectral operators only holds, however, if $\bbV^{-1}$ is used as GFT and the GSO at hand is diagonalizable. 

{Generalizations of graph filters were introduced within the class of \textit{linear} graph-aware signal operators.} These include node-variant  \cite{segarra2017optimal}  and edge-variant graph filters \cite{coutino2019advances}, whose expressions are respectively given by
\begin{eqnarray}\label{E:Filter_node_an_edge_variant}
\bbH_{nv}:=\sum_{l=0}^{L-1}\diag(\bbh_l) \mathbf{S}^l\;\;\;\textrm{and}\;\;\;\bbH_{ev}:=\sum_{l=0}^{L-1}(\bbH_l \circ \bbS) \bbS^{l-1},
\end{eqnarray}
where $\circ$ denotes the Hadamard product, $\bbh_l$ is a vector of dimension $N$, and $\bbH_l$ is a sparse matrix with the same support than $\bbS$. {Compared with its (node-invariant) counterpart in \eqref{E:Filter_input_output_vertex}, we observe that the output generated by a node-variant filter can also be viewed as a linear combination of locally shifted inputs $\bbx^{(l)}=\bbS^l\bbx$, but in this case each node has the flexibility of using a different set of weights. The flexibility is even larger for edge-variant graph filters, since nodes can change the weight they give to \textit{each} of their neighbors (cf. \textit{all} of their neighbors for node-variant filters). } Since both $\bbH_{nv}$ and $\bbH_{ev}$ build on a polynomial definition, they can seamlessly operate over digraphs.  They thus inherit most of the properties described for the original polynomial graph filters in \eqref{E:Filter_input_output_vertex}.

\subsection{Graph neural network architectures}

{Graph filters have also been used to define nonlinear operators that account for the topology of the graph,} such as median filters \cite{segarra2016center} and Volterra graph filters. All these works build their definitions from the polynomial expression in \eqref{E:Filter_input_output_vertex} and, hence, can handle digraphs, although some of their properties (e.g., the conditions that a signal needs to satisfy to be a root of a median graph filter \cite{segarra2016center}) require minor modifications. A case of particular interest is that of  deep graph neural network (NN) architectures \cite{gama2018convolutional}, which have attracted significant attention in recent years {to tackle machine learning problems involving network data. Traditional (e.g., convolutional) NNs have been remarkably successful in tasks involving images, video and speech, all of which represent data with an underlying Euclidean domain that is regularly sampled over a grid-like structure. However, said structure one almost takes for granted is missing when it comes to signals defined on graphs. As argued next, GSP offers an ideal framework to fill in this fundamental gap.} 

The overall idea in graph NN architectures is to define an input-output relation by using a concatenation of $L_N$  layers composed of a linear transformation that combines the different signal values and a scalar (point-wise) nonlinear function that increases the expressiveness of the mapping. Mathematically, with $\bbx$ and $\bby$ denoting the input and the output to the overall NN architecture and $\ell$ being the layer index, we have that  
\begin{align}
\label{E:Generic_Graph_NN_architecture_input}
\bbz^{(0)}&=\bbx,\hspace{.2cm}\text{and}\hspace{.2cm}\bby = \bbz^{(L_N)},\hspace{.2cm}\text{where}\hspace{.2cm}\\
\hbz^{(\ell)}&= \ccalT_{\bbtheta^{(\ell)}}^{(\ell)} \Big\{\bbz^{(\ell-1)}\Big| \ccalG \Big\},\;\;1\leq \ell \leq L_N,\\
\bbz^{(\ell+1)}_{ij} &= \sigma^{(\ell)}_{\ccalG} \Big([\hbz^{(\ell)}]_{ij}\Big),\;\;1\leq \ell \leq L_N\; \text{and}\; \text{all}\;i,j\in\ccalN. \label{E:Generic_Graph_NN_architecture_output} 
\end{align}
In the expressions above, $\bbz^{(\ell)}$ is the output of layer $\ell$ and serves as input to layer $\ell+1$. The transformation $\ccalT_{\bbtheta^{(\ell)}}^{(\ell)}\{\cdot|\ccalG\}$ is the linear operator implemented at layer $\ell$, $\bbtheta^{(\ell)}$ are the learnable parameters that define such a transformation, and $\sigma^{(\ell)}_{\ccalG}: \reals \rightarrow \reals$ is a scalar nonlinear operator (possibly different per layer). When applied to graph signals, the NN architecture in \eqref{E:Generic_Graph_NN_architecture_input}-\eqref{E:Generic_Graph_NN_architecture_output} must account for the topology of the graph and, for that reason, the dependence of both the linear and nonlinear operators on $\ccalG$ was made explicit. In most works, the role of the graph is considered when defining the linear operator in $\ccalT_{\bbtheta^{(\ell)}}^{(\ell)}\{\cdot|\ccalG\}$ with the most widely-used approach for graph convolutional NNs being to replace $\ccalT_{\bbtheta^{(\ell)}}^{(\ell)}\{\cdot|\ccalG\}$ with a graph filter. {Precisely in inspiring this approach is where GSP insights and advances have been transformative, since basic shift-invariance properties and convolution operations are otherwise not well defined for graph signals~\cite{bronstein2017geometric}.}

Early contributions following the graph filtering rationale emerged from the machine learning community. The spectral approach in~\cite{bronstein2017geometric} relies on the Laplacian eigenvectors $\bbV$ and parametrizes the transformation $\ccalT_{\bbtheta^{(\ell)}}^{(\ell)}\{\cdot|\ccalG\}=\bbV \diag(\bbtheta^{(\ell)})\bbV^T $ via $\tbg=\bbtheta^{(\ell)}$, the filter's frequency response in \eqref{E:Filter_input_output_frequency} that is learned using backpropagation. While successful in many applications, when dealing with digraphs these approaches suffer from the same limitations as those discussed for their linear counterparts. Moreover, scalability is often an issue due to the computational burden associated with calculating the eigenvectors of large (albeit sparse) graphs. Alternative architectures proposed replacing $\ccalT_{\bbtheta^{(\ell)}}^{(\ell)}\{\bbz^{(\ell-1)}|\ccalG\}$ with $(\bbI-\theta^{(\ell)} \bbA)\bbz^{(\ell-1)}$, where $\bbA$ is the (possibly non-symmetric) adjacency matrix of the graph and $\theta^{(\ell)}$ is a learnable scalar. To increase the number of parameters some authors have considered learning the non-zero entries of $\bbA$, assuming that its support is known. A more natural approach is to replace $\ccalT_{\bbtheta^{(\ell)}}^{(\ell)}\{\cdot|\ccalG\}$ with the polynomial filter in \eqref{E:Filter_input_output_vertex} and consider the filter taps $\bbh=\bbtheta^{(\ell)}$ as the parameters to be learned; see \cite{gama2018convolutional} and references therein. Once again, implementing \eqref{E:Generic_Graph_NN_architecture_input}-\eqref{E:Generic_Graph_NN_architecture_output} with $\bbH^{(\ell)}=\sum_{l=0}^{L_\ell-1}h_l^{(\ell)} \bbS^l$ in lieu of $\ccalT_{\bbtheta^{(\ell)}}^{(\ell)}\{\cdot|\ccalG\}$ exhibits a number of advantages since: (i) the graph filter is always well-defined (even for non-diagonalizable GSOs); (ii) the degree of the filter controls the complexity of the architecture (number of learnable parameters); and (iii) the polynomial definition guarantees that the resultant graph filter can be implemented efficiently (via the successive application of sparse matrices), which is essential in scaling to large datasets. {As in standard NN architectures, graph NN parameters (i.e., the filter coefficients for each of the layers) are learned using stochastic gradient descent. For supervised learning tasks, the goal is to minimize a suitable loss function over a training set of (labeled) examples. The sparsity of $\bbS$ and the efficient implementation of polynomial graph filters [cf. (iii)] are cardinal properties to keep the overall computational complexity in check.}

Beyond graph \textit{convolutional} NNs, the aforementioned findings are also valid for \textit{recurrent} graph NNs. Furthermore, one can also replace the graph filter $\bbH^{(\ell)}=\sum_{l=0}^{L_\ell-1}h_l^{(\ell)} \bbS^l$ {either with a set of parallel filters, or with its node-variant $\bbH_{nv}^{(\ell)}$ or edge-variant $\bbH_{ev}^{(\ell)}$ counterparts. All of them preserve the distributed implementation of \eqref{E:Filter_input_output_vertex} while increasing the number of learnable parameters.} As a result, the use of polynomial-based graph filter definitions that operate directly in the nodal domain to design NN architectures for digraphs {opens a number of research avenues for deep learning over digraphs; see, e.g., \cite{bronstein2017geometric,gama2018convolutional,wu2020comprehensive} as well as other relevant papers in this special issue for additional details.}



\section{Inverse problems on digraphs}\label{S:inverse_probs}

Inverse problems such as sampling and deconvolution played a central role in the development of GSP.
Different modeling assumptions must be considered when addressing these problems for digraphs; a good practice is to leverage the concepts introduced in Sections~\ref{S:GFT_and_Dics_for_Digraphs} and~\ref{S:Filters_for_Digraphs} and balancing practical utility with mathematical tractability. {For instance, parsimonious signal models based on graph smoothness or bandlimitedness are widely adopted. Alternatively, observations can be modeled as the outputs of graph filters driven by white, sparse, or piece-wise constant inputs. This approach is particularly useful in applications dealing with diffusion processes defined over real-world networks with directional links.} In this section, we formally introduce a selection of prominent inverse problems, present established approaches for their solution, and identify the main challenges when the signals at hand are defined over digraphs. 

\subsection{{Sampling and reconstruction}}
\label{ss:sampling}

The sampling of graph signals and their subsequent reconstruction have arguably been the most widely-studied problems within GSP \cite[Ch. 9]{djuric2018cooperative}. Broadly speaking, the objective is to infer the value of the signal at every node from the observations at a few nodes by leveraging the structure of the graph. To describe the problem formally, let us introduce the fat, \textit{binary}, $M \times N$ sampling matrix $\bbC_\ccalM$ and define the sampled signal as $\bar{\bbx} = \bbC_\ccalM \bbx$. Notice that if $\ccalM$ represents the subset of $M<N$ nodes where the signal is sampled, $\bbC_\ccalM$ has exactly one nonzero element per row, and the position of those non-zero elements correspond to the indexes of the nodes in $\ccalM$, then the signal $\bar{\bbx}$ is indeed a selection of $M$ out of the $N$ elements of $\bbx$. {This raises two fundamental questions, namely how to reconstruct $\bbx$ from $\bar{\bbx}$ and how to design $\bbC_\ccalM$ to facilitate this reconstruction.}

Starting with the first question, early works assumed the graph to be undirected and the signal $\bbx$ to be bandlimited, i.e., to be a linear combination of just a few leading eigenvectors of the GSO. The GSO was typically set to the Laplacian $\bbL$, with its eigenvectors $\bbV=[\bbv_1,...,\bbv_N]$ being real-valued and orthogonal. That is, the signal was assumed to be expressible as $\bbx=\sum_{k=1}^K \tilde{x}_k\bbv_k:=\bbV_K \tbx_K$, where $\tbx_K\in\reals^K$ collects the $K$ active frequency coefficients and $\bbV_K$ is a submatrix of the GFT.  Indeed, since the leading eigenvectors in $\bbV$ are  those with the smallest total variation [cf.~{\eqref{eqn_TV_general}}], this model was originally motivated by the practical importance of signals that vary smoothly with the underlying graph. Under the bandlimited assumption, the sampled signal $\bar{\bbx}$ is  given by $\bar{\bbx} = \bbC_\ccalM \bbx = \bbC_\ccalM \bbV_K \tilde{\bbx}_K$. Clearly, if the linear transformation represented by matrix $\bbC_\ccalM \bbV_K\in \reals^{M\times K}$ is full column rank (that is, if $\bbC_\ccalM \bbV_K$ has rank $K$), then $\tilde{\bbx}_K$ can be recovered from $\bar{\bbx}$. Once the coefficients $\tilde{\bbx}_K$ are known, the signal in the original domain can be found as $\bbx = \bbV_K \tilde{\bbx}_K = \bbV_K  (\bbC_\ccalM \bbV_K)^{\dagger}  \bar{\bbx}$. Hence, the critical factor to characterize the recovery of $\bbx$ from $\bar{\bbx}$ is the invertibility (and conditioning) of matrix $\bbC_\ccalM \bbV_K$, which is a submatrix of $\bbV$ formed by the $K$ columns corresponding to the active frequencies and the $M$ rows corresponding to the sampled nodes in $\ccalM$. Notice that a key difference with sampling in classical SP is that designing matrix $\bbC_\ccalM$ as a regular sampler is meaningless in GSP since the node indexing is completely arbitrary. 
Indeed, multiple approaches have been proposed to identify the most informative nodes on a graph for subsequent reconstruction. This is tantamount to leveraging the (spectral) properties of $\bbC_\ccalM \bbV_K$ in order to design sampling matrices $\bbC_\ccalM$ that lead to an optimal reconstruction. {For example by maximizing the minimum singular value of $\bbC_\ccalM \bbV_K$, the sampling set is designed to minimize the effect of noise in a mean-squared error sense \cite{chen_sampling_2015}.}
Equally important, the fact that the reconstruction matrix is a submatrix of the eigenvectors of the graph has also been exploited to design optimal low-pass graph filtering operators that can reconstruct the original signal $\bbx$ by implementing local exchanges \cite{segarra2015reconstruction}, as well as efficient algorithms that leverage the sparsity of the graph to compute $\bbV_K$ efficiently \cite{le2017approximate}.

When dealing with the sampling and reconstruction of signals defined on digraphs, a number of challenges arise. 
As introduced in Section~\ref{Ss:GFT_for_Digraphs}, multiple definitions of GFT coexist for digraphs. 
Some of those are based on generalizations of smoothness and lead to real-valued orthogonal dictionaries. 
In those cases, the results presented for signals in undirected graphs still hold, but the connections with polynomial low-pass filtering and the ability to find the eigenvectors efficiently are lost.  Alternatively, one can use (a subset of) the eigenvectors of the non-symmetric GSO as the basis for the signal $\bbx$. {The caveats, in this case, being that the GSO needs to be diagonalizable and that the resulting eigenvectors $\bbV$ are neither orthogonal nor real-valued. The latter point implies that the frequency coefficients $\tbx_K$ are complex-valued as well, so that the recovery methods for digraphs must be conceived in the complex field.} Regarding the loss of orthogonality, this will typically deteriorate the conditioning of the submatrix $\bbC_\ccalM \bbV_K$, which is critical in regimes where noise is present and $M$ is close to $K$. 
Hence, when dealing with the sampling of real-world signals defined over digraphs, a first step is to decide which type of signal dictionary is going to be used. This likely depends on the prior domain knowledge as well as on the properties of the signals at hand.
If no prior knowledge exists, schemes considering different dictionaries (at the expense of increasing the sample complexity) may be prudent. 
Moreover, in the cases where the selected basis is composed of the eigenvectors of the GSO, the recovery problems need to be formulated in the complex domain and oversampling is likely to be required in scenarios where noise, outliers, or model mismatches are present.

Additional models for the observed signal have been studied in the digraph literature, including the cases where (i)~the $K$ dictionary atoms spanning $\bbx$ are not known a priori (thus leading to a sparse regression problem)~\cite{chen_sampling_2015}; 
(ii)~the observations do not correspond to values of $\bbx$ but rather of $\bbS^i \bbx$ for varying $i$ (which can be interpreted as sampling an evolving network process as opposed to a static one)~\cite{marques_2016_sampling}; (iii)~total variation metrics are considered in the form of regularizers or constraints~\cite{chen_2015_signal}; and (iv)~the signal $\bbx$ is modeled as the output of a graph filter excited by a structured input~\cite{segarra2017blind, iglesias_2018_demixing}.
We will revisit the two last cases while studying the next collection of inverse problems.

\subsection{(Blind) deconvolution, system identification, and source localization}

We now introduce a family of recovery and reconstruction problems involving signals over digraphs. 
The common denominator across all of them is the assumption that the generative model $\bby=\bbH\bbx$ holds, where $\bby$ is a (partially) observed graph signal, $\bbH$ is a \textit{linear graph filter}, and $\bbx$ is a potentially unknown and structured input. 
Building on this model and assuming that we have access to samples of the output $\bby$, the supporting digraph, and side information on $\bbH$ and $\bbx$, the goal is to recover (i)~the graph filter $\bbH$ (system identification); (ii)~the values of $\bbx$ (deconvolution); (iii)~the support of $\bbx$ (source localization); or (iv)~both the graph filter and the values of $\bbx$ (blind deconvolution). 
Since graph filters can be efficiently used to model local diffusion dynamics, the relevance of the mentioned schemes goes beyond signal reconstruction and permeates to broader domains such as opinion formation and source identification in social networks, inverse problems of biological signals supported on graphs, and modeling and estimation of diffusion processes in multi-agent networks, {all of which are typically directed. In particular, we envision applications in marketing where, e.g., social media advertisers want to identify a small set of influencers so that an online campaign can go viral; in healthcare policy implementing network analytics to infer hidden needle-sharing networks of injecting drug users; or, in environmental monitoring using wireless sensor networks to localize heat or seismic sources.}
As an encompassing formal framework, consider the following optimization problem
\begin{align}\label{E:optim_decon_general}
\{ \bbx^*, \bbh^*, \bby^* \} = \argmin_{\{ \bbx, \bbh, \bby\}} \,\, \ccalL_0 \left( \bby, \sum_{l=0}^{L-1}h_l \mathbf{S}^l \bbx \right) + \alpha_x r_x(\bbx) \nonumber\\
+ \alpha_h r_h(\bbh) + \alpha_y r_y(\bby), \quad \text{s. to} \,\, \bbx \in \mathcal{X}, \bbh \in \mathcal{H}, \bby \in \mathcal{Y}; 
\end{align}
where $\ccalL_0$ is a loss function between the observed signal $\bby$ and its prediction generated by the chosen $\bbx$ and $\bbh$. The regularizers $r_x$, $r_h$, and $r_y$ promote desirable features on the optimization variables, and $\mathcal{X}$, $\mathcal{H}$ and $\mathcal{Y}$ represent pre-specified feasibility sets. While for the undirected case the generative filter $\bbH$ can be either defined in the spectral or in the vertex domain, in \eqref{E:optim_decon_general} the polynomial form has been selected. As pointed out in Section \ref{S:Filters_for_Digraphs}, the reasons for this choice are multiple: polynomial filters are always well defined (even for non-diagonalizable GSOs); the number of parameters is $L$ ({in contrast with $N$ for those spectral formulations that do not consider an explicit parametrization}), which is beneficial in the context of inverse problems; and the filter can be used to capture distributed diffusion dynamics on directed networks, strengthening the practical value of the formulation in \eqref{E:optim_decon_general}.  
Finally, even though~\eqref{E:optim_decon_general} was posited for the generic case where both $\bbx$ and $\bbh$ are unknown and $\bby$ might be only partially observed, it is immediate to incorporate perfect knowledge of any of these variables, just by fixing its value and dropping the corresponding feasibility constraint and regularization~term. 

Focusing first on the problem of deconvolution, notice that the non-symmetric filter $\bbH$ is completely known since both the GSO $\bbS$ and the filter coefficients $\bbh$ are assumed to be given.
The goal is then to use incomplete observations of $\bby$ to recover the values of $\bby$ in the non-observed nodes and to obtain the seeding values in $\bbx$.
Leveraging the notation introduced in Section~\ref{ss:sampling}, we denote by $\bar{\bby} = \bbC_\ccalM \bby = \bbH_\ccalM \bbx$ the sampled output, with $ \bbH_\ccalM = \bbC_\ccalM \bbH$ being the corresponding $M$ rows of $\bbH$.
Since $\bby$ and $\bbx$ are graph signals of the same size, the deconvolution problem is ill-posed when $M < N$. 
Hence, to overcome this we may assume some structural prior on the input $\bbx$. 
A common assumption is that $\bbx$ is sparse. This corresponds to setups where the observed signal $\bby$ can be accurately modeled by a few sources percolating across the entire network. 
Applications fitting this setup range from social networks where a rumor originated by a small group of people is spread across the network via local opinion exchanges, to brain networks where an epileptic seizure emanating from few regions is later diffused across the entire brain \cite{segarra2015reconstruction}.
Formally, problem \eqref{E:optim_decon_general} reduces to 
\begin{equation}\label{E:optim_decon}
\bbx^* = \argmin_{\bbx} \,\, \|\bar{\bby} - \bbH_\ccalM \bbx \|_2^2 + \alpha_x \| \bbx \|_1,
\end{equation}
which is a classical sparse-regression problem, with well-established results showing that the recovery performance provably depends on the coherence of the non-symmetric matrix $\bbH_\ccalM$.
The $\ell_1$-norm regularizer in \eqref{E:optim_decon} acts as a convex surrogate of the sparsity-measuring $\ell_0$ pseudo-norm.
Whenever sparsity is assumed as a structural property of the input and the emphasis is on recovering the support of $\bbx$, \eqref{E:optim_decon} and variations thereof (with imperfect knowledge of $\bbh$) are referred to as source localization problems. In terms of the samples of $\bby$ that are observed, the optimal selection (in cases where this selection can be designed) is non-trivial and considerations similar to those discussed in Section~\ref{ss:sampling} apply here as well. Finally, note that the generative model $\bby = \bbH \bbx$ can also be used for undirected graphs and, as a result, the formulation in \eqref{E:optim_decon} and the associated algorithms can be used in such a case, the main difference being that the theoretical analysis of identifiability and recovery is simpler when the GSO (and hence the filter) is symmetric. 

Moving on to the system identification problem, where the main objective is to find the filter coefficients $\bbh$, it is crucial to note that $\bby$ is a bilinear function in $\bbh$ and $\bbx$. 
Hence, if we assume that $\bbx$ is given, then the system identification problem is very similar to the deconvolution problem where the roles of $\bbx$ and $\bbh$ are interchanged.
In terms of structural priors for an unknown $\bbh$, sparsity can also be employed. 
More specifically, it is instrumental to consider a weighted $\ell_1$-norm regularization $r_h(\bbh) = \| \diag(\bbomega) \bbh \|_1$, where $\bbomega\in \reals_+^{L}$ is a weighting vector whose weights increase with $l=1,...,L$, the entry index.
In this way, coefficients associated with higher powers of $\bbS$ in the filter specification are more heavily penalized, thus promoting a low-complexity and numerically-stable model for explaining the observed data. For undirected graphs, the cost $\ccalL_0$ that enforces the generative graph filter model to hold is oftentimes formulated in the spectral domain, bypassing the need of computing the powers of $\bbS$. While we advocate working on the nodal domain, when the GSO is diagonalizable, formulating the problem in the spectral domain is also feasible for the directed case. The matrices mapping the unknown $\bbh$ to the observations $\bar{\bby}$ would be complex-valued, but the optimization would still be carried over the real-valued vector $\bbh$. From an algorithmic perspective, the main challenge would be to find the eigenvectors of the non-symmetric $\bbS$, while from the analytical point of view the issue would be the characterization of the conditioning of the (complex-valued) matrix that maps $\bbh$ to $\bar{\bby}$.

The more challenging problem of blind deconvolution arises when both the input $\bbx$ and the filter coefficients $\bbh$ are unknown. 
To formally tackle this problem, we explicitly write the fact that $\bby$ is a bilinear function of $\bbh$ and $\bbx$ as $\bby = \ccalA (\bbx \bbh^T)$, where the linear operator $\ccalA$ is a function of the non-symmetric $\bbS$ and acts on the outer product of the sought vectors.
A direct implementation of the general framework~\eqref{E:optim_decon_general} can be employed for the problem of blind identification where the goodness-of-fit loss $\| \bby - \ccalA (\bbx \bbh^T)\|_2^2$ is combined with structure-promoting regularizers for both $\bbx$ and $\bbh$. Notice, however, that this leads to a non-convex optimization problem for which alternating minimization schemes (e.g., a block coordinate descent method that alternates between $\bbx$ and $\bbh$) can be implemented.
In order to derive a convex relaxation, notice that $\bby$ is a linear function of the entries of the rank-one matrix $\bbZ = \bbx \bbh^T$. This motivates the statement of the following convex optimization problem
\begin{equation}\label{E:optim_decon_blind}
\bbZ^* = \argmin_{\bbZ} \,\, \|\bby - \ccalA(\bbZ) \|_2^2 + \alpha_1 \| \bbZ \|_* + \alpha_2 \| \bbZ \|_{2,1}.
\end{equation}
The nuclear norm regularizer $\| \cdot \|_*$ in \eqref{E:optim_decon_blind} promotes a low-rank solution since we know that $\bbZ$ should be the outer product of the true variables of interest $\bbx$ and $\bbh$. 
On the other hand, the $\ell_{2,1}$ mixed norm $\| \bbZ \|_{2,1} = \sum_{i=1}^N \| \bbz_i \|_2$ is the sum of the $\ell_2$-norms of the rows of $\bbZ$, thus promoting a row-sparse structure in $\bbZ$. 
This is aligned with a sparse input $\bbx$ forcing rows of $\bbZ$ to be entirely zero from the outer product.
After solving for $\bbZ^*$, one may recover $\bbx$ and $\bbh$ from, e.g., a rank-one decomposition of $\bbZ^*$.

\begin{figure*}
	\centering\includegraphics[width=0.85\linewidth]{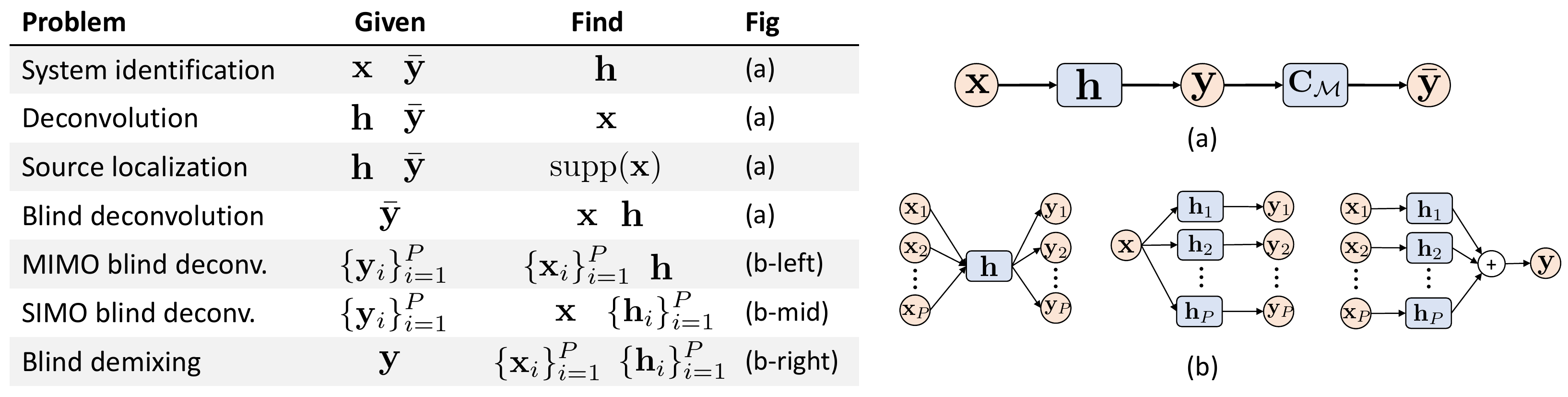}
	\vspace{-4mm}
	\caption{Summary of the inverse problems introduced. In the schematic representation, graph signals are depicted as red circles and graph operators as blue rectangles. 
		Notice that filters are functions of the coefficients $\bbh$ and the GSO $\bbS$ [cf.~\ref{E:Filter_input_output_vertex}]. 
		However, since $\bbS$ is assumed to be known for every problem considered, we succinctly represent filters by their coefficients $\bbh$. 
		The first four problems refer to the single input, single output scenario with blind deconvolution being the most challenging since only (a sampled version of) the output is observed. Notice that the problem frameworks in {(b)} can be further extended to the case where the output is partially observed as in (a). We omit this illustration to minimize redundancy and because these more challenging problems are generally ill-posed even in the case where $\bby$ is fully observed.}
	\vspace{-3mm}
	\label{F:deconvolution}
\end{figure*}

Extensions to multiple input-output pairs (with a common filter) along with theoretical guarantees for the case where the GSO $\bbS$ is normal (i.e., $\bbS \bbS^H = \bbS^H \bbS$) can be found in~\cite{segarra2017blind}. 
Interestingly, it was empirically observed and theoretically demonstrated that blind deconvolution in circulant graphs (such as the directed cycle that represents the domain of classic SP) corresponds to the most favorable setting.
The related case of a single graph signal as the input to multiple filters (generating multiple outputs) was recently studied in~\cite{zhu_2019_estimating}, thus providing a generalization of the classical blind multi-channel identification problem in digital SP. 
Moreover, \cite{iglesias_2018_demixing} addresses the blind demixing case where a single observation formed by the sum of multiple outputs is available, and it is assumed that these outputs are generated by different sparse inputs diffused through different graph filters.
This variation of the problem is severely ill-posed and strong regularization conditions should be assumed to ensure {recovery, with the problem being easier if the graph filters are defined over different GSOs $\{\bbS_i\}_{i=1}^I$.}
Fig.~\ref{F:deconvolution} provides an overarching view of the problems mentioned in this section.
See also \cite{chen_2015_signal} for additional signal-recovery problems that can  be written in the encompassing framework of~\eqref{E:optim_decon_general}.

As previously explained, a key feature that allows using the problem formulations introduced in this section for signals defined on digraphs is that the generative graph filter was incorporated in polynomial form. Unfortunately, many of the \textit{theoretical guarantees} for solving these problems heavily rely on the spectral analysis of the GSO, thus assuming symmetry or at least normality of $\bbS$.
One of the main remaining challenges for inverse problems in digraphs is the derivation of recovery guarantees along with the identification of key performance drivers that can accommodate non-diagonalizable GSOs and generalized (complex) eigenvectors.
Another important potential research direction is the incorporation of alternative generative models by replacing graph filters with the more general graph-signal operators presented in Section~\ref{S:Filters_for_Digraphs}, such as node-variant and edge-variant filters~\eqref{E:Filter_node_an_edge_variant} or graph neural networks~\eqref{E:Generic_Graph_NN_architecture_input}-\eqref{E:Generic_Graph_NN_architecture_output}.
Especially in this latter case, system identification and blind deconvolution would become extremely challenging due to the incorporation of nonlinearities, making the convex relaxation in \eqref{E:optim_decon_blind} based on the linear operator $\ccalA$ no longer valid.


\section{Statistical digraph signal processing}\label{S:Satistical_SP}

Randomness is pervasive in engineering and graph signals are not an exception. For this reason, here we build on the results presented in the previous sections to discuss recent advances and challenges to develop statistical models for \textit{random} graph signals defined over digraphs. In the field of statistics, graphs quickly emerged as a convenient intuitive mathematical structure to describe complex statistical dependencies across multi-dimensional variables. A prominent example is given by Markov random fields (MRFs), which are symmetric graphical models whose edges capture conditional dependencies across the variables represented by the nodes. Inference over MRFs is computationally affordable and, for the particular case of the signals being Gaussian, the graph describing the MRF can be inferred directly from the precision (inverse covariance) matrix of the data. In parallel, directed graphs have been used to capture one-directional conditional dependence (hence causal) relations, with Bayesian networks -- which on top of being directed are acyclic -- being the most tractable graphical model within this class.   

The GSP literature has also contributed to the statistical modeling of random graph signals. {The first step is to postulate how the graph structure plays a role towards shaping the signal's statistical properties and, then, analyze how the model put forth can be used to tackle inference tasks more effectively.} As in the case of graphical models, most existing results focused on undirected graphs. {Arguably, the most relevant line of work has been the generalization of the definition of weak stationarity to signals supported either on undirected graphs or on graphs whose GSO is a normal matrix~\cite[Ch. 12]{djuric2018cooperative}.
While this latter characterization includes some directed graphs (such as circulant and skew-Hermitian), the definitions cannot be applied to a generic non-symmetric GSO. The key contribution of the papers reviewed in~\cite[Ch. 12]{djuric2018cooperative}} was to provide a dual definition for stationary graph processes which was consistent with the vertex and frequency interpretations of graph signals. 
Specifically, it was stated that a zero-mean random graph signal $\bbx$ was weakly stationary on a known graph $\ccalG$ if: (i) its covariance matrix has the same eigenvectors than those of the GSO; or, equivalently, (ii) the process can be modeled as the output of a graph filter excited with a white input. This allowed establishing parallelisms with the classical definition of weak stationarity for time-varying signals and opened the door to the development of efficient algorithms that estimate the second moment of a graph stationary process using less samples (e.g., if the eigenvectors of the covariance and the GSO are the same, instead of estimating the $N^2$ entries of the covariance matrix, one can focus on estimating only its $N$ eigenvalues). 

However, from the initial discussion in Section \ref{S:Filters_for_Digraphs}  it follows that this convenient equivalence between the frequency and the vertex domains does not hold for digraphs. Indeed, if the GSO is not a normal matrix its eigenvectors cannot coincide with those of the covariance matrix, which is guaranteed to be normal. As a result, one must adapt the definitions and sacrifice some of the properties shown for the symmetric case. To be mathematically precise, let us recall that $\bbx$ is a zero-mean random process defined on the directed graph $\ccalG$ with GSO $\bbS$, and let us denote by $\bbC_\bbx:=\mathbb{E}[\bbx\bbx^T]$ the $N\times N$ covariance matrix of $\bbx$. We say that the random graph signal $\bbx$ is stationary in the non-symmetric$\,$$\bbS$$\,$if$\,$it$\,$can$\,$be$\,$described$\,$as
\begin{eqnarray}\label{E:Stationary_graph_process_directed}
\bbx=\bbH\bbw,\;\;\textrm{with}\;\;\mathbf{H}:=\sum_{l=0}^{L-1}h_l \mathbf{S}^l\;\textrm{and}\;\mathbb{E}[\bbw\bbw^T]=\bbI,
\end{eqnarray}
where $L\leq N$ and $\bbw$ is a white zero-mean random signal. By adopting the generative model in \eqref{E:Stationary_graph_process_directed}, it follows that the covariance of $\bbx$ can be written as $\bbC_\bbx=\mathbb{E}[\bbx\bbx^T]=\bbH \mathbb{E}[\bbw\bbw^T] \bbH^T=\bbH\bbH^T$, which is not a polynomial on $\bbS$, but on both $\bbS$ and $\bbS^T$. As a result, it is no longer true that $\bbC_\bbx$ is diagonalized by the GFT associated with $\bbS$. Nonetheless, the model in \eqref{E:Stationary_graph_process_directed} is still extremely useful since it: (i) provides an intuitive explanation of the notion of graph stationarity; (ii) can be used to establish connections with higher-order auto-regressive directed structural equation models in statistics; and (iii) gives rise to efficient estimators that, rather than targeting the estimation of the full covariance, try to estimate the \textit{filter coefficients} $\bbh$. Indeed, this latter point is also relevant for undirected graphs. While approaches that focus on the spectral definition of stationary processes require estimating the $N$ eigenvalues of $\bbC_\bbx$ (i.e., the power spectral density of the process $\bbx$), the generative approaches based on \eqref{E:Stationary_graph_process_directed} open the door to imposing additional structure to the generative filter (e.g., considering an FIR/IIR filter with a number of coefficients $L$ much smaller than the number of nodes $N$), resulting in considerable gains in terms of either the sampling complexity or the estimation error.  

The generative model in \eqref{E:Stationary_graph_process_directed} can be generalized or constrained to fit a range of suitable scenarios. Focusing first on the input signal, cases of practical interest include: (i) considering non-white input processes $\bbw$ with known covariance; (ii) requiring $\bbw$ not only to be white but also independent; and (iii) particularizing the distribution of $\bbw$ to tractable and practically meaningful cases. Two examples that fall into the last category are modeling $\bbw$ as either a Gaussian or a (signed) Bernoulli vector, which is particularly relevant in the context of diffusion of sparse signals. Alternatively, the model in \eqref{E:Stationary_graph_process_directed} can be enlarged by considering other linear graph-signal operators as generators, including the node-variant and edge-variant graph filters discussed in \eqref{E:Filter_node_an_edge_variant}. Recent works have also proposed nonlinear generative models that exploit results in the deep learning literature to generate random signals over directed and undirected graphs. For example, one can take the architecture in \eqref{E:Generic_Graph_NN_architecture_input}-\eqref{E:Generic_Graph_NN_architecture_output}, replace  $\ccalT_{\bbtheta^{(\ell)}}^{(\ell)}\{\cdot|\ccalG\}$ with $\bbH^{(\ell)}=\sum_{l=0}^{L_\ell-1}h_l^{(\ell)} \bbS^l$, use a random realization of the white signal $\bbw$ as input, and then view the output of the graph NN architecture as the random process to be modeled. While characterizing how the coefficients $\{\bbh^{(\ell)}\}_{\ell=1}^{L_N}$ affect the statistical properties of the output is certainly relevant, equally interesting problems arise when the goal is to use a set of realizations of the output $\bbx$ to learn the parameters of the nonlinear generative model (i.e., the filter coefficients $\{\bbh^{(\ell)}\}_{\ell=1}^{L_N}$ ) that best fit the available observations.

The statistical models briefly reviewed in the previous paragraphs accounted for non-symmetric interactions among variables and can be leveraged, for example, to enhance covariance estimation schemes, to denoise a set of graph signals observations, or to interpolate (predict) values of graph signals using as input observations at only a subset of nodes. Maybe less obvious but arguably equally important, the postulated models can also be used to infer the graph itself. Indeed, if one has access to a set $\ccalX:=\{\bbx_r\}_{r=1}^R$ of $R$ realizations of $\bbx$ and the graph is sufficiently sparse (so that the number of edges $|\ccalE|$ is much smaller than $N^2$), one could identify the $L+|\ccalE|$ degrees of freedom in \eqref{E:Stationary_graph_process_directed} from the $RN$ values in $\ccalX$, provided that $R$ is sufficiently large. This is partially the subject of the next section, which deals with the problem of inferring the topology of digraphs from a set of nodal observations.

\section{Digraph topology inference}\label{S:topo_id}

Capitalizing on the GSP advances surveyed so far requires a specification of the underlying digraph. However, $\ccalG$ is often unobservable and, accordingly, network topology inference from a set of (graph-signal) measurements is a prominent yet challenging problem, {even more so when the graph at hand is directed.} Early foundational contributions can be traced back several decades to the statistical literature of graphical model selection; see e.g.,~\cite[Ch. 7]{kolaczyk2009book} and the opening of Section \ref{S:Satistical_SP}. Discovering directional influence among variables is at the heart of causal inference, and identification of  cause-effect digraphs (so-termed structural causal models) from observational data is a notoriously difficult problem~\cite[Chs. 7 and 10]{Peters2017}.  Recently, the fresh modeling and signal representation perspectives offered by GSP have sparked renewed interest in the field. Initial efforts have mostly focused on learning undirected graphs, which naturally give rise to more tractable (and often uniquely identifiable) formulations~\cite{SI_SPMAG2019}. {Therefore, in the sequel we outline a} few noteworthy digraph topology identification approaches that are relevant to (or are informed by) the GSP theme of this paper. In accordance with our narrative's leitmotif, we emphasize the key differences with the undirected case and review the main challenges associated with the new formulations.

We initiate our exposition with structural equation modeling, which broadly encapsulates a family of statistical methods that
describe causal relationships between interacting variables in a complex system. This is pursued through the estimation
of linear relationships among endogenous as well as exogenous traits. Structural equation models (SEMs) have been extensively
adopted in economics, psychometrics, social sciences, and genetics, among other disciplines; see e.g.,~\cite{ggtopoid2018piee}. 
SEMs postulate a linear time-invariant network model {of the following form, where the GSO is specified as the adjacency matrix $\bbS = \bbA$},
\begin{equation}\label{E:SEM}
x_{it}=\!\!\sum_{j=1, j\neq i}^N {S}_{ij}x_{jt}+\omega_{ii}u_{it}+\epsilon_{it}, \: i\in \ccalN \:\: \Rightarrow \bbx_t={\bbS}\bbx_t+\mathbf{\Omega}\bbu_t+\bbepsilon_t,
\end{equation}
where $\bbx_t=[x_{1t},\ldots,x_{Nt}]^T$ represents a graph signal of endogenous variables at discrete time $t$ and $\bbu_t=[u_{1t},\ldots,u_{Nt}]^T$ is a vector of exogenous influences. The term ${\bbS}\bbx_t$ in \eqref{E:SEM} models network effects, implying $x_{it}$ is a linear combination of the instantaneous values $x_{jt}$ of node $i$'s in-neighbors $j\in\ccalN_i$. The signal $x_{it}$ also depends on $u_{it}$, where weight $\omega_{ii}$ captures the level of influence of external sources and we defined $\mathbf{\Omega}:=\textrm{diag}(\omega_{11},\ldots,\omega_{NN})$. Vector $\bbepsilon_t$ represents measurement errors and unmodeled dynamics.  Depending on the context, $\bbx_t$ can be thought of as an output signal while $\bbu_t$ corresponds to the excitation or control input. In the absence of noise and letting $\mathbf{\Omega}=\bbI$ for simplicity, \eqref{E:SEM} becomes $\bbx_t=\bbH\bbu_t$, where $\bbH:=(\bbI-{\bbS})^{-1}$ is a polynomial graph filter as in \eqref{E:Filter_input_output_vertex}. 

Given snapshot observations $\ccalX:=\{\bbx_t,\bbu_t\}_{t=1}^T$, SEM parameters {$\bbS$} and $\bbomega:=[\omega_{11},\ldots,\omega_{NN}]^T$ are typically estimated via penalized least squares, for instance by solving
\begin{eqnarray}\label{E:SEM_estimation}
{\hat{\bbS}} &=& \argmin_{{\bbS},\bbomega}{}\;\;\sum_{t=1}^T\|\bbx_t-{\bbS}\bbx_t+\mathbf{\Omega}\bbu_t\|_2^2+\alpha\|{\bbS}\|_1,\hspace{1cm} \nonumber\\
&~&\textrm{ s. to } \:\:\mathbf{\Omega}=\textrm{diag}(\bbomega),\quad {S}_{ii}= 0,\: i=1,\ldots,N,
\end{eqnarray}
where the $\ell_1$-norm penalty promotes sparsity in the adjacency matrix. Both edge sparsity and endogenous inputs play a critical role in guaranteeing that the SEM parameters \eqref{E:SEM} are uniquely identifiable; see also~\cite{ggtopoid2018piee}. Acknowledging the limitations
of linear models,~\cite{Shen2017kernel_SEM} leverages kernels within the SEM framework to model nonlinear pairwise dependencies
among network nodes; see Section~\ref{Ss:gene_networks} for results on the identification of gene-regulatory networks.

While SEMs only capture contemporaneous relationships among the nodal variables (i.e., SEMs are memoryless), sparse vector autoregressive models (SVARMs) account for linear time-lagged (causal) influences instead; see e.g.,~\cite{varm_group_sparse}. Specifically, for a given model order $L$ and unknown sparse evolution matrices $\{{\bbS}^{(l)}\}_{l=1}^L$, SVARMs postulate a multivariate linear dynamical model of the form
%
$\bbx_t=\sum_{l=1}^L{\bbS}^{(l)}\bbx_{t-l}+\bbepsilon_t$.
%
Here a directed edge from vertex $j$ to $i$ is typically said to be present in $\ccalG$ if ${S}_{ij}^{(l)}\neq 0$ for all $l=1,\ldots,L$. The aforementioned AND rule is often explicitly imposed as a constraint during estimation of SVARM parameters, through the requirement that all matrices ${\bbS}^{(l)}$ have a common support. This can be achieved for instance via a group lasso penalty, that promotes sparsity over edgewise coefficients ${\bbs}_{ij}:=[{S}_{ij}^{(1)},\ldots,{S}_{ij}^{(L)}]^T$ jointly~\cite{varm_group_sparse}. The sparsity assumption is often well-justified due to physical considerations or for the sake of interpretability, but here (as well as with SEMs) it is also critical to reliably estimate $\ccalG$ from limited and noisy time-series data $\ccalX:=\{\bbx_t\}_{t=1}^T$.


SVARMs are also central to popular digraph topology identification approaches based on the principle of Granger causality; see e.g.,~\cite[Ch. 10]{Peters2017}. Said principle is based on the concept of precedence and predictability,
where node $j$'s time series is said to ``Granger-cause'' the time series at node $i$ if knowledge of $\{x_{j,t-l}\}_{l=1}^L$
improves the prediction of $x_{it}$ compared to using only $\{x_{i,t-l}\}_{l=1}^L$. Such a form of causal dependence defines the status of a candidate edge from $j$ to $i$, and it can be assessed via judicious hypothesis testing~\cite{ggtopoid2018piee}.  Recently, a notion different from Granger's was advocated to associate a graph with causal network effects among vertex time series, effectively blending VARMs with graph filter-based dynamical models. The so-termed causal graph process (CGP) introduced in~\cite{mei_tsp2017} {also considers $\bbS = \bbA$ and} has the form 
\begin{eqnarray}\label{E:causal_graph_process}
\bbx_t&=& \sum_{l=1}^L\sum_{i=0}^l h_{li}{\bbS}^i\bbx_{t-l}+\bbepsilon_t
=(h_{10}\bbI +h_{11}{\bbS})\bbx_{t-1}+\ldots \nonumber\\
&+&(h_{L0}\bbI+\ldots+h_{LL}{\bbS}^L)\bbx_{t-L}+\bbepsilon_t,
\end{eqnarray}
where ${\bbS}$ is the (possibly asymmetric) adjacency matrix encoding the unknown graph topology. The CGP model corresponds to a generalized VARM with coefficients given by $\bbH_l({\bbS},\bar{\bbh}):=\sum_{i=0}^l h_{li}{\bbS}^i$  where  $\bar{\bbh}\!:=\![h_{10}, h_{11},\ldots,h_{li},\ldots h_{LL}]^T\!$. This way, the model can possibly account for multi-hop nodal influences per time step. Unlike SVARMs, matrices $\bbH_l({\bbS},\bar{\bbh})$ need not be sparse for larger values of $l$, even if ${\bbS}$ is itself sparse.  Given data $\ccalX:=\{\bbx_t\}_{t=1}^T$ and a prescribed value of $L$,  to estimate ${\bbS}$ one solves the non-convex optimization problem
\begin{equation}\label{E:mei_graph_estimation}
{\hbS} = \argmin_{{\bbS},\bar{\bbh}}\;\;\sum_{t=L+1}^{T}\Big\|\bbx_t-\sum_{l=1}^L\bbH_l({\bbS},\bar{\bbh})\bbx_{t-l}\Big\|^2+\alpha\|{\bbS}\|_1+\beta\|\bar{\bbh}\|_1.
\end{equation}
Similar to sparse SEMs in \eqref{E:SEM_estimation} and SVARMs, the estimator encourages sparse graph topologies. Moreover, the $\ell_1$-norm regularization on the filter coefficients $\bar{\bbh}$ effectively implements a form of model-order selection. A divide-and-conquer heuristic is advocated in~\cite{mei_tsp2017} to tackle the challenging problem \eqref{E:mei_graph_estimation}, whereby one: (i) identifies the filters $\bbH_l:=\bbH_l({\bbS},\bar{\bbh})$ so that $\bbx_t \approx \sum_{l=1}^L\sum_{i=0}^l \bbH_l\bbx_{t-l}$, exploiting that $\bbH_l$ and $\bbH_{l'}$ commute for all $l,l'$; (ii) recovers a sparse ${\bbS}$ using the estimates $\{\hbH_l\}$ and leveraging the shift-invariant property of graph filters; and (iii) estimates $\bar{\bbh}$ given $\{\hbH_l,{\hbS}\}$ via the lasso. For full algorithmic details and accompanying convergence analysis, please see~\cite{mei_tsp2017}. 

In~\cite{topoid_directed_dsw18} observations from $M$ network processes are modeled as the outputs of a polynomial graph filter [i.e, $\bbx_m=\bbH\bbw_m$ as in \eqref{E:Filter_input_output_vertex}], excited by (unobservable) zero-mean independent graph  signals $\bbw_m$ with arbitrarily-correlated nodal components. Observations of the output signals along with prior statistical information on the inputs are first utilized to identify the non-symmetric diffusion filter $\bbH$. Such problem entails solving a system of quadratic matrix equations, which can be recast as a smooth quadratic minimization subject to Stiefel manifold constraints; see~\cite{topoid_directed_dsw18} for details.  Given an estimate $\hbH$, the approach in~\cite{topoid_directed_dsw18} to infer the digraph topology is to find a {generic} GSO $\bbS$ that satisfies certain desirable topological properties and commutes with $\bbH$. For instance, focusing on the recovery of sparse graphs one solves
\begin{equation}\label{E:general_problem}
\hat{\bbS} = \argmin_{\bbS} \
\|\bbS\|_1 ,   \quad                       
\text{s.~to }\:
\bbS \in \ccalS,  \quad
\| \hbH \bbS - \bbS \hbH \|_{F} \leq \eps,
\end{equation}
where  $\ccalS$ is a convex set specifying the type of GSO sought (say, the adjacency matrix of a digraph), and the constraint $\| \hbH \bbS - \bbS \hbH \|_{F} \leq \eps$  encourages the filter $\bbH$ to be a polynomial in $\bbS$ while accounting for estimation errors; see \cite{mei_tsp2017,djuric2018cooperative,SI_SPMAG2019}. Imposing this last constraint offers an important departure from related (undirected) graph learning algorithms in~\cite{SI_TSIPN2017,SI_SPMAG2019},~\cite[Ch. 13]{djuric2018cooperative}, which identify the structure of network diffusion processes from observations of stationary signals [cf. \eqref{E:Stationary_graph_process_directed} but with symmetric $\bbH$]. These approaches first estimate the \textit{eigenvectors} of $\bbH$, and then constrain $\bbS$ to be diagonalized by those eigenvectors in a convex problem to recover the unknown eigenvalues. While this naturally entails a search over the lower-dimensional space of GSO eigenvalues, the formulation \eqref{E:general_problem} avoids computing an eigendecomposition and, more importantly, solving a problem over complex-valued variables. This was not an issue in~\cite{SI_TSIPN2017}, since the focus therein was on undirected graphs with real-valued spectrum. In closing, note that the graph filtering model advocated in~\cite{topoid_directed_dsw18}  is a special case of \eqref{E:causal_graph_process}  {provided that $\bbS=\bbA$}, and instead of multivariate  time-series data one relies on independent replicates from multiple network processes (obtained e.g., via interventions as in causal inference~\cite{Peters2017}).

\begin{figure*}
	\centering\includegraphics[width=0.90\linewidth]{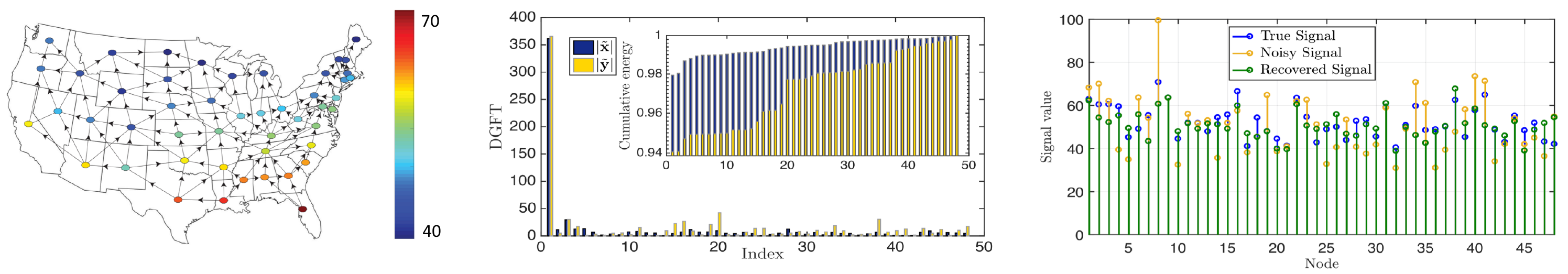}
	\vspace{-3mm}
	\caption{Temperature denoising using the DGFT~\cite{RAGE_GFT_TSP_2019}.
		(Left)~Graph signal of average annual temperature in Fahrenheit for the contiguous US states. In the depicted digraph, a directed edge joins two states if they share a border, and the edge directions go from South to North. 
		(Center)~DGFT of the original signal ($\tilde{\bbx}$) and the noisy signal ($\tilde{\bby}$), along with their cumulative energy distribution across frequencies. (Right)~A sample realization of the true, noisy, and recovered temperature signals for a filter bandwidth $K = 3$.}
	\vspace{-3mm}
	\label{F:temp_usa}
\end{figure*}

\section{Applications}\label{S:Apps}

We highlight four real-world applications of the methods surveyed in this paper.
The experiments were chosen to demonstrate the practical value of SP schemes applied to digraphs, with the diversity of the datasets considered (climate records, text excerpts, handwritten characters, and gene expression levels) underscoring the versatility of the~tools.

\vspace{.2cm}
\noindent\textbf{Frequency analysis for temperature signal denoising.} 
We consider a digraph $\ccalG$ of the $N=48$ contiguous United States (Alaska and Hawaii are excluded). 
A directed edge joins two states if they share a border, and the edge direction is set so that the state whose barycenter is more to the South points to the one more to the North.
As the graph signal $\bbx\in\reals^{48}$ we consider the average annual temperature of each state; see Fig.~\ref{F:temp_usa} (left).
The temperature map confirms that the latitude affects the average temperatures of the states, justifying the proposed latitude-based graph-construction scheme.

We determine a GFT basis $\bbU$ for this digraph via spectral dispersion minimization [cf.~\eqref{eq:delta_opt_prob}] and test its utility in a denoising task.
More specifically, our goal is to recover the temperature signal from noisy measurements $\bby = \bbx {+} \bbw_\bby$, where the additive noise $\bbw_\bby$ is a zero-mean, Gaussian random vector with covariance matrix $10 \bbI_{N}$.
To achieve this, we implement a low-pass graph filter that retains the first $K$ components of the signal's DGFT and eliminates the rest, i.e., $\tilde{\bbh}=[\tilde{h}_{1},\ldots,\tilde{h}_{N}]^T$, where $\tilde{h}_{k} = \ind{k \leq K}$ and $K$ is a prescribed spectral window size.
Hence, we estimate the true temperature signal as $\hat{\bbx} = \bbU \text{diag}(\tilde{\bbh})  \tilde{\bby} = \bbU \text{diag}(\tilde{\bbh}) \bbU^T \bby$.


The original signal $\bbx$ is bandlimited compared to the noisy signal $\bby$, which spans a broader range of frequencies; see Fig.~\ref{F:temp_usa} (center).
To better observe the low-pass property of $\bbx$, we also plot the cumulative energy of both $\bbx$ and $\bby$, defined by the percentage of the total energy present in the first $k$ frequency components for $k=1,\dots,N$. 
Setting the spectral window at $K=3$, the average recovery error $e_f = \| \hat{\bbx} - \bbx \|/\|\bbx\|$ over $1000$ Monte-Carlo simulations of independent noise was of approximately $12\%$. 
Fig.~\ref{F:temp_usa} (right) shows a realization of the noisy graph signal $\bby$ superimposed with the denoised temperature profile $\hat{\bbx}$ and it can be seen that, indeed, $\hat{\bbx}$ closely approximates $\bbx$.
{The recovery error increases when the edge directions are ignored (i.e., $\ccalG$ is treated as undirected) and when they are selected randomly (i.e., every edge is directed but the specific orientation is chosen uniformly at random between the two possibilities), as opposed to following the South to North orientation that captures the temperate flow}; see~\cite{RAGE_GFT_TSP_2019} for additional details and experiments.

\begin{figure}
	\centering\includegraphics[width=0.99\linewidth]{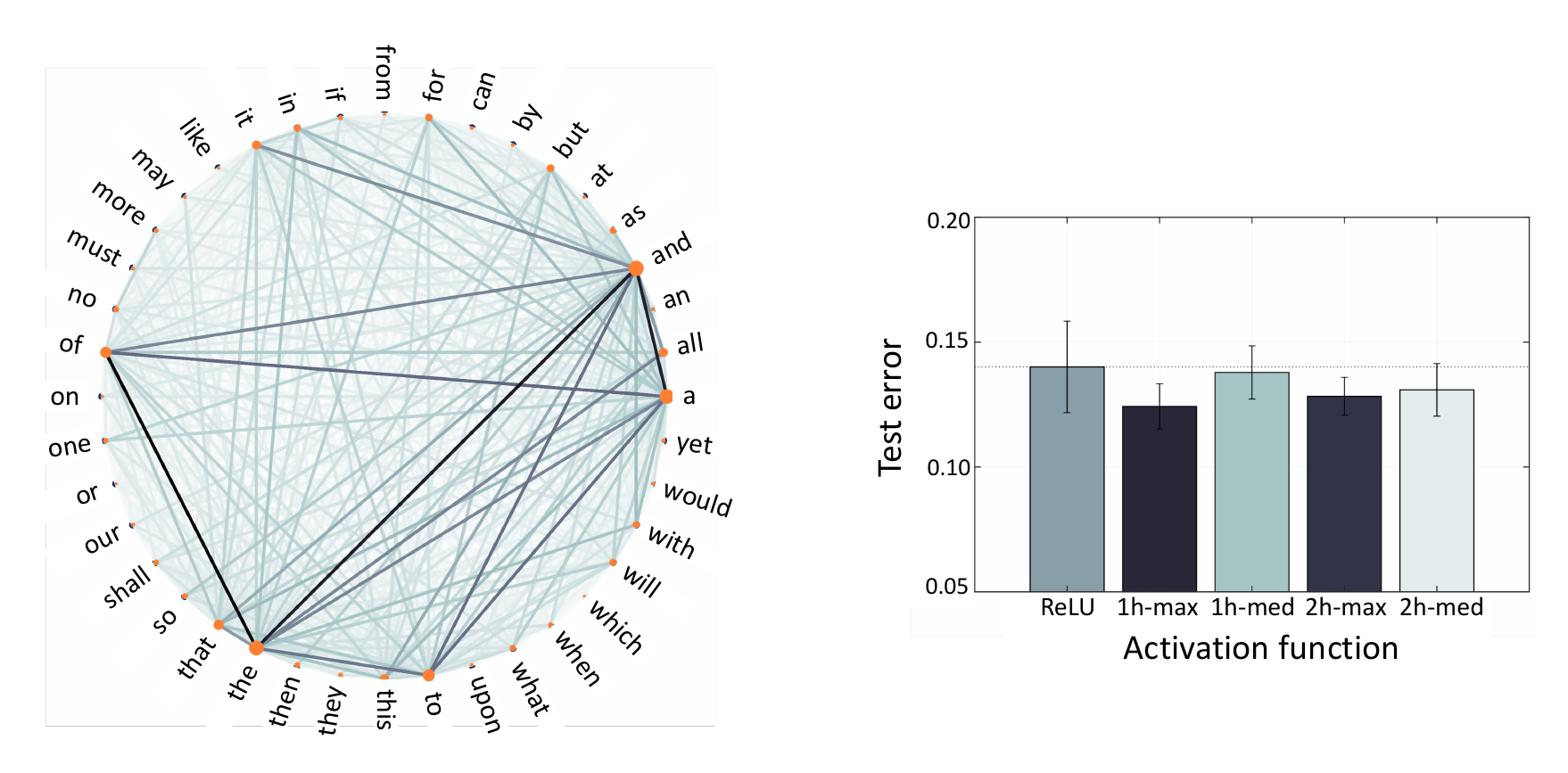}
	\vspace{-3mm}
	\caption{Identifying the author of a text using GNNs \cite{ruiz2020invariance}.
		(Left)~Example of a WAN with 40 function words as nodes built from the play ``The Humorous Lieutenant'' by John Fletcher. The radius of the nodes is proportional to the word count and the darker the edge color, the higher the edge weight. Directionality has been ignored for ease of representation, but the graph NNs are defined on the directed WAN.
		(Right)~Authorship attribution test error in graph NN architectures with localized activation functions for the classification of Emily Br\"onte versus her contemporaries.}
	\vspace{-3mm}
	\label{F:wan}
\end{figure}

\vspace{.2cm}
\noindent\textbf{Graph neural networks for authorship attribution.} 
We illustrate the performance of graph NNs for classification in digraphs through an authorship attribution problem
based on real data. The goal is, using a short text excerpt as input, to decide whether the text was written by a particular author. To capture the style of an author, we consider author-specific word adjacency networks (WANs), which are digraphs whose nodes are function words (i.e., prepositions, pronouns, conjunctions and other words with syntactic importance but little semantic meaning~\cite{ruiz2020invariance}) and whose edges represent probabilities of directed co-appearance of two function words within texts written by the author; see Fig.~\ref{F:wan} (left).

We select $N = 211$ functions words as nodes and build the WAN for Emily Br\"onte. 
More specifically, we count the number of times each pair of function words co-appear in $10$-word windows, while also recording their relative order. 
We then normalize the counts out of each node to sum up to one, thus obtaining a weighted digraph whose weights are between $0$ and $1$.
As for the graph signals, they are defined as each function word's count among $1,000$ words.  Splitting Emily Br\"onte's texts between training and test sets on an $80$--$20$ ratio, her WAN is generated from function
word co-appearance counts in the training set only. 
The graph signals in the training set correspond to $1000$-word excerpts by Br\"onte and by a pool of other 21 contemporary authors. Each graph signals has an associated binary label where $1$ indicates that the text has been written by Br\"onte and excerpts by the rest of the authors are labeled as $0$.
Test samples are defined analogously. 
The training and test sets consisted of $1,092$ and $272$ excerpts, both with equally balanced classes, and cross-entropy was chosen as the loss function.

Several specific graph NN architectures were compared in this experiment, all of them following the general structure in~\eqref{E:Generic_Graph_NN_architecture_input}-\eqref{E:Generic_Graph_NN_architecture_output} but for different choices of the nonlinearity $\sigma^{(\ell)}_{\ccalG}$. 
Indeed, the popular pointwise ReLU was contrasted with more sophisticated graph-localized (but not necessarily pointwise) median and maximum activation functions; see \cite{ruiz2020invariance} for details.
Fig.~\ref{F:wan} (right) presents the authorship attribution accuracy results after conducting ten rounds of simulations by varying the training and test splits.
We can see that median and max graph NNs did consistently better than the ReLU graph NNs on discerning between texts written by Br\"onte and other authors in the pool. {Localized activation functions outperform the pointwise ReLU, with smaller average test errors as well as smaller deviations around this average}. {Equally important, the simulations also show that their associated error is 1-2\% lower than that achieved by NN architectures that symmetrize the WAN.} This superior performance underscores the importance of leveraging the directed graph structure in the architecture of graph NNs, not only in the linear operators via the incorporation of graph filters but also in the determination of nonlinearities.

\begin{figure}
	\centering\includegraphics[width=0.75\linewidth]{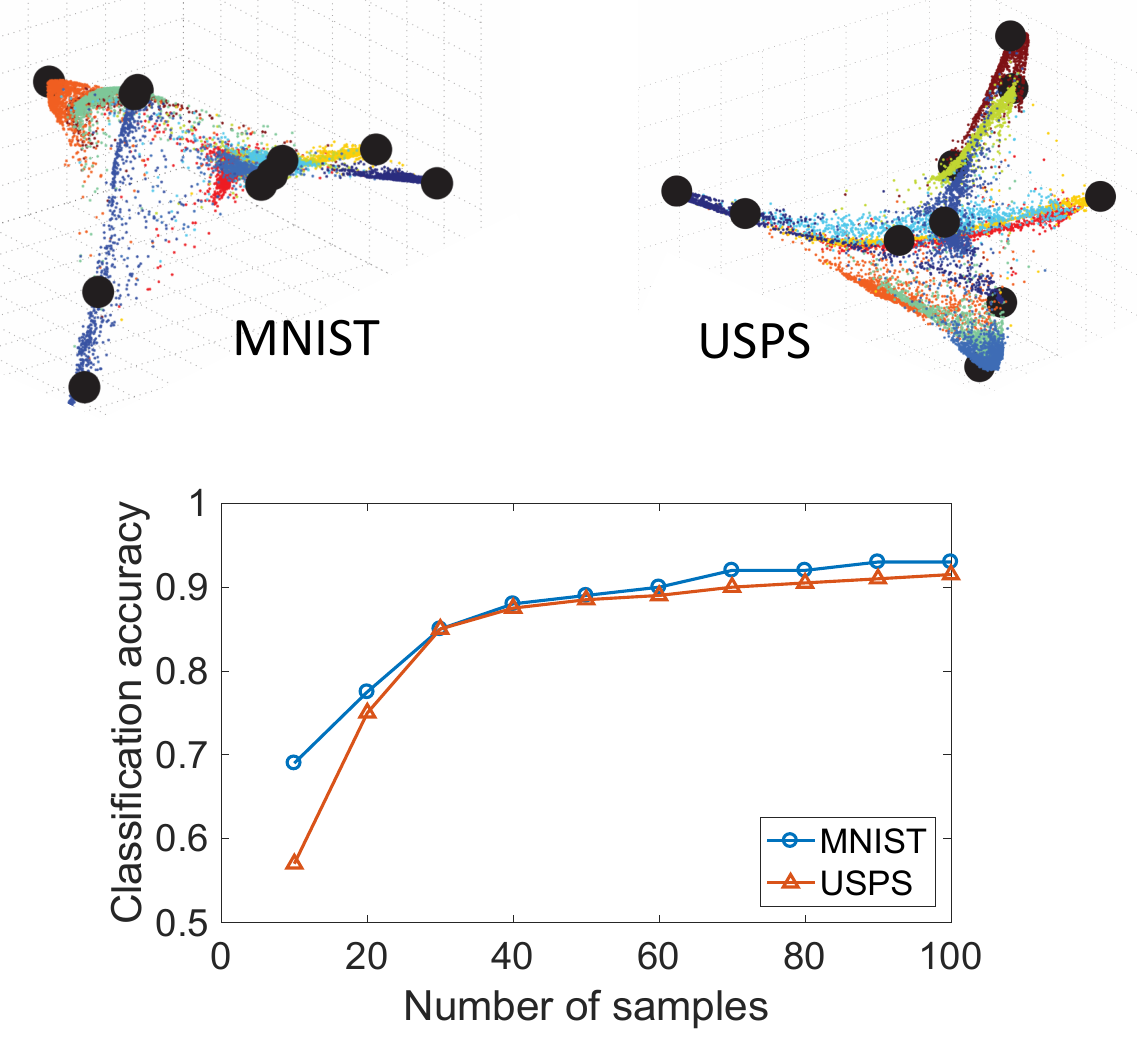}
	\vspace{-3mm}
	\caption{Semi-supervised learning for handwritten digit classification via sampling of graph signals~\cite{chen_sampling_2015}. 
		(Top left)~A three-dimensional representation of the MNIST images colored by true class (digits 0 through 9). The ten enlarged nodes correspond to the identified optimal samples. 
		(Top right)~Analogous of the left plot, but for the USPS dataset. 
		(Bottom)~Classification accuracy as a function of the number of samples used for interpolation for both~datasets.}
	\vspace{-3mm}
	\label{F:mnist}
\end{figure}

\vspace{.2cm}
\noindent\textbf{Graph sampling for handwritten digit recognition.} 
Our goal here is to employ the sampling theory introduced in Section~\ref{ss:sampling} to classify handwritten digits with minimal labels as developed in~\cite{chen_sampling_2015}.
More precisely, we consider a directed graph whose $N = 10,000$ nodes correspond to gray-scaled images in the MNIST dataset equally distributed among the ten classes ($0$-$9$ digit characters).
The edges are obtained from a $12$-nearest neighbor construction computed from the Euclidean distance between vector representations of the images. 
The graph is directed by construction, since one node being in the $12$-nearest neighborhood of another node does not guarantee that the relation in the opposite direction holds.
{This directionality can be especially relevant in the treatment of outliers in the embedded space, where every outlier
still has an incoming neighborhood of size $12$ but does not belong to the incoming neighborhood of other nodes, thus having a minimal effect in the label propagation.} 	
The edges are then weighted using a normalized Gaussian kernel so that, within each neighborhood of size $12$, the closer connections have a larger weight.
Intuitively, images representing the same digit tend to have similar pixel values and, hence, are more likely to belong to the neighborhood of each other.
Thus, if we consider the value of the signal at a given node to be the digit represented by the image associated with that node, the whole graph signal will be piece-wise constant in the graph and thus amenable to being reconstructed from observations at a few nodes.
Furthermore, to account for the fact that the signal values are categorical, instead of considering a graph signal of dimension $\bbx \in \reals ^N$, we consider the alternative binary matrix representation $\bbX \in \reals^{N \times 10}$, where $X_{ij} = +1$ if the $i$-th image is a picture of the digit $j$ and $X_{ij} = -1$ otherwise.
Each column of $\bbX$ is modeled as a bandlimited signal that can be written as the linear combination of the $K$ leading columns of $\bbV$, the eigenvectors of the non-symmetric adjacency matrix $\bbA$.

The graph representation of the MNIST digits is shown in Fig.~\ref{F:mnist} (left), where the edges were removed for clarity and the coordinates of each node are given by the corresponding rows of the first three columns of the iGFT $\bbV$. The enlarged black nodes indicate the optimal choice for $10$ samples.
Optimality, in this case, refers to the design of $\bbC_\ccalM$ to maximize the minimum singular value of $\bbC_\ccalM \bbV_K$ (cf. Section~\ref{ss:sampling}). Given that we have to (pseudo-)invert this matrix for reconstruction, a good condition number entails a robust behavior in the presence of noise.
It is apparent that the images representing the same digit form clusters and that the optimal samples boil down to choosing representative samples from each cluster.
The same procedure can be repeated for the USPS handwritten digits dataset consisting of $N = 11,000$ images, to obtain Fig.~\ref{F:mnist} (center).
For both cases, one can compute the classification accuracy obtained from the reconstructed graph signals for a different number of optimal samples; Fig.~\ref{F:mnist} (right). 
As expected, the accuracy increases with the number of samples. 
Furthermore, note that even when observing only 50 samples ($0.5\%$ of the dataset for the case of MNIST and $0.45\%$ for USPS), the reconstruction accuracy is almost $0.9$, highlighting the importance of incorporating the graph structure via optimal samplers that can accommodate directed graphs. 
This method was shown to outperform other graph-based active semi-supervised learning techniques; see~\cite{chen_sampling_2015} for additional details and experiments.

\vspace{.2cm}
\noindent\textbf{Kernel-based topology inference for gene expression data.}
\label{Ss:gene_networks} Consider now the problem of identifying gene regulatory topologies, where nodes represent individual genes and directed edges encode causal regulatory relationships between gene pairs.
{Due to the inherent directional nature of regulatory interactions~\cite[Ch. 7.3]{kolaczyk2009book}, we must recover a digraph as opposed to an undirected relational structure. In this context,}
we compare the inferred digraphs recovered when implementing different kernels for SEM inference. 
The experiments were performed on gene regulatory data collected from $T=69$ unrelated Nigerian individuals, under the International HapMap project; see~\cite{Shen2017kernel_SEM} and references therein for additional details. 
From the 929 identified genes, expression levels and the genotypes of the expression quantitative trait loci (eQTLs) of $N=39$ immune-related genes were selected and normalized.
Genotypes of eQTLs were considered as exogenous inputs $\bbu_t$ whereas gene expression levels were treated as the endogenous variables $\bbx_t$ [cf.~\eqref{E:SEM}].

\begin{figure}
	\centering\includegraphics[width=0.80\linewidth]{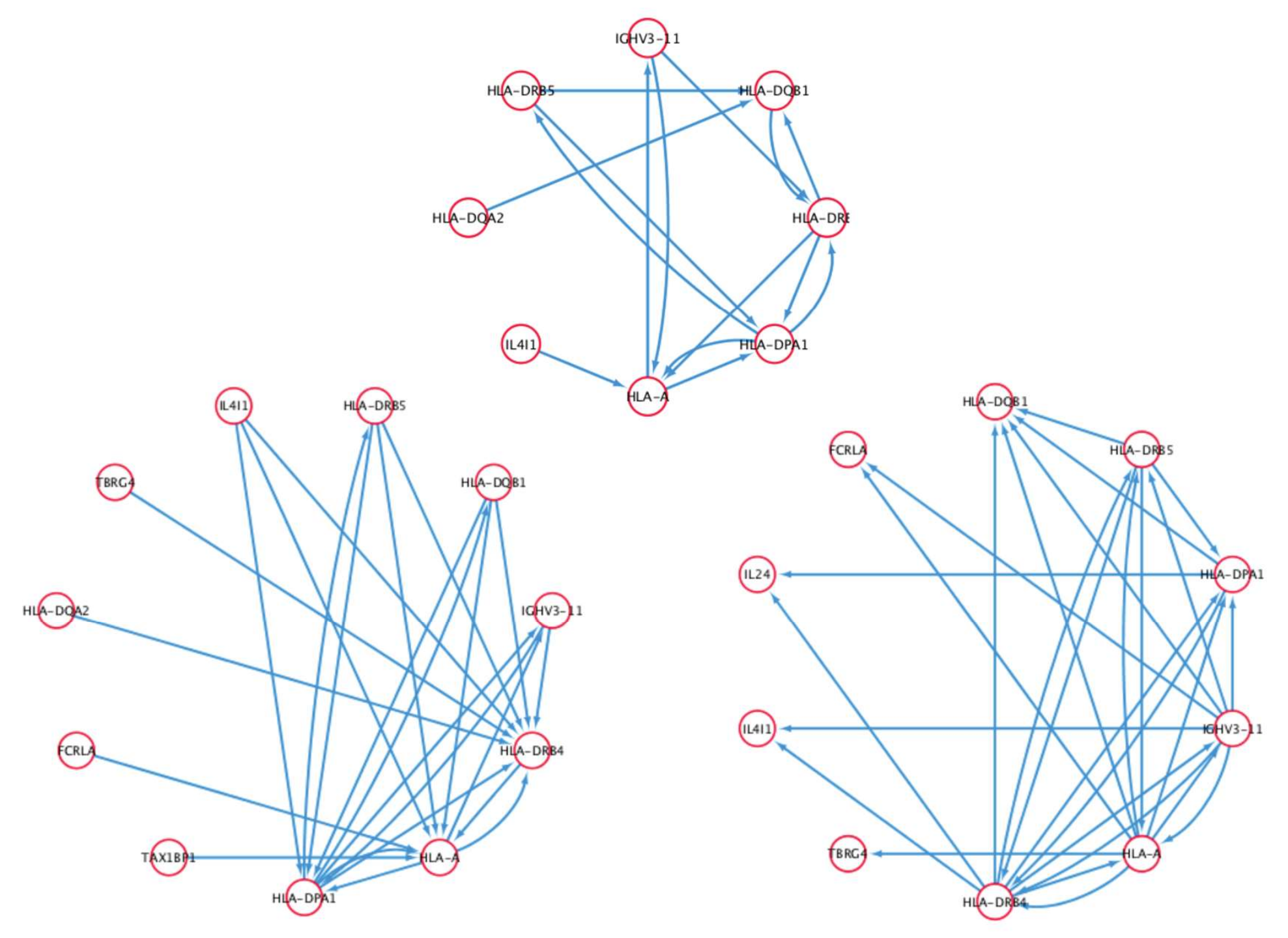}
	\vspace{-3mm}
	\caption{Inferring (directed) gene regulatory networks from expression data~\cite{Shen2017kernel_SEM}. The networks are inferred following the SEM formulation in~\eqref{E:SEM_estimation} for a linear kernel (top), a polynomial kernel of order 2 (bottom left), and a Gaussian kernel with unit variance (bottom right).}
	\vspace{-3mm}
	\label{F:topo_id}
\end{figure}

Fig.~\ref{F:topo_id} depicts the identified topologies, where the different graphs correspond to different choices for the kernel and the visualizations only include nodes that have at least a single incoming or outgoing edge. 
More precisely, Fig.~\ref{F:topo_id} (left) portrays the resulting network based on a linear SEM while the center and right panels in Fig.~\ref{F:topo_id} illustrate the results from nonlinear SEMs based on a polynomial kernel of second order and a Gaussian kernel with unit variance, respectively. 
In the three cases, the identified networks are very sparse, and the nonlinear approaches unveil all edges identified by the linear SEMs, alongside with a number of additional edges. 
Clearly, considering the possibility that interactions among genes may be driven by nonlinear dynamics, nonlinear frameworks encompass linear approaches and facilitate the discovery of causal (directed) patterns not captured by linear SEMs. {The newly unveiled gene regulatory interactions could potentially be the subject of further studies and direct experimental corroboration by geneticists to improve our understanding of causal influences among immune-related genes across humans}.

\section{Emerging topic areas and conclusions}\label{S:Conclusions}

Contending that signals defined on digraphs are of paramount practical importance, this paper outlined recent approaches to model, process, and learn from these graph signals.
Accordingly, this tutorial stretched in a comprehensive and unifying manner all the way from the definition of graph Fourier transforms and graph-signal operators especially designed for digraphs to the problem of inferring the digraph itself from the observed signals.
A wide range of signal recovery problems was selectively covered, focusing on inverse problems in digraphs including sampling, deconvolution, and system identification.
A statistical viewpoint for signal modeling was also discussed by extending the definition of weak stationarity of random graph processes to the directed domain. The last stop was to review recent results that applied the tools surveyed in this tutorial to the problem of learning the topology of a digraph from nodal observations, an approach that can lead to meaningful connections between GSP and the field of causal inference in statistics.   
A common theme in the extension of established GSP concepts to the less explored realm of digraphs is that definitions and notions that heavily rely on spectral properties are challenging to generalize whereas those that can be explicitly postulated in the vertex domain are more amenable to be extended to digraphs.

A diverse gamut of potential research avenues naturally follows from the developments presented.
Efficient approaches for the computation of the multiple GFTs for digraphs (akin to the fast Fourier transform in classical SP) would facilitate the adoption of this methodology in large-scale settings.
The incorporation of nonlinear (median, Volterra, NNs) graph signal operators as generative models for the solution of inverse problems is another broad area of promising research.
Deep generative models for signals defined in regular domains (such as images) have shown remarkable success over the last years, and part of that success can be extended to our more challenging domain.
Equally interesting, the use of deep learning to generate the graphs themselves (as opposed to the graph signals) is recently gaining traction so that, along the lines in this tutorial, one can conceive neural network architectures that learn (and even generate) digraphs from training graph signals while encoding desirable topological features. 
One last direction of future research is the extension of the concepts here discussed to the case of higher-order directed relational structures.
The generalization of GSP to hypergraphs through tensor models and simplicial complexes has been explored in recent years, but their analysis in directed scenarios is almost uncharted research territory.

%

\bibliographystyle{IEEEtran}
%
\bibliography{citations}

\end{document}